\documentclass[conference,letterpaper, onecolumn]{IEEEtran}

\usepackage[utf8]{inputenc} 
\usepackage[T1]{fontenc}
\usepackage{url}
\usepackage{ifthen}
\usepackage{cite}
\usepackage[cmex10]{amsmath} 

\usepackage{graphicx}
\usepackage{amsfonts}
\usepackage{amssymb}
\usepackage{epsfig}	
\usepackage{epstopdf}
\usepackage{fancyhdr} 
\usepackage{mathrsfs}
\usepackage{mathtools}
\usepackage{caption}
\usepackage{subcaption}
\usepackage{float}
\usepackage{xcolor}
\usepackage{multirow}
\usepackage{adjustbox}
\usepackage{colortbl}
\usepackage{tabulary}
\usepackage{etoolbox}
\usepackage{multicol, multirow}
\usepackage{booktabs}
\usepackage{pdfpages}
\usepackage{nomencl}
\usepackage{adjustbox}
\usepackage{notoccite}

\usepackage{setspace}
\usepackage{algorithmic}
\usepackage{algorithm}
\usepackage{array}
\usepackage{textcomp}
\usepackage{stfloats}
\usepackage{verbatim}
\usepackage{graphicx}
\usepackage{enumerate}
\usepackage{eucal}
\usepackage{comment}
\usepackage{xcolor}
\usepackage{adjustbox}

\usepackage{tikz}
\usepackage{color, colortbl}
\usepackage{nicematrix}

\usepackage{float}

\usepackage{bigdelim}
\usepackage{arydshln}
\usepackage{mathrsfs}
\usepackage{mathtools}

\newcommand{\BB}{\mathbb{B}}

\newcommand{\JJ}{\mathpzc{J}}

\newcommand{\II}{\mathpzc{I}}

\newcommand{\F}{\mathcal{F}}

\newcommand{\Z}{\mathbb{Z}}

\newcommand{\ZZ}{\mathcal{Z}}

\newcommand{\Prob}{\text{Pr}}

\newtheorem{lemma}{Lemma}
\newtheorem{theorem}{Theorem}

\newtheorem{corollary}{Corollary}

\newtheorem{defn}{Definition}

\DeclareMathAlphabet{\mathpzc}{OT1}{pzc}{m}{it}

\newcommand{\neqlvl}{\eta}

\newcommand{\nodelab}{\mathfrak{l}}
\newcommand{\nodemat}{\mathfrak{m}}
\newcommand{\nodeshft}{\mathfrak{t}}
\newcommand{\nodepol}{\mathfrak{Q}}
\newcommand{\nodeskew}{\mathpzc{s}}

\newcommand{\TT}{\mathcal{T}}
\renewcommand{\vec}[1]{\underline{#1}}

\begin{document}
\title{Numerical Stability of DFT Computation for Signals with Structured Support} 


  
  
\author{%
  \IEEEauthorblockN{Charantej Reddy Pochimireddy}
  \IEEEauthorblockA{Email: ee18resch01010@iith.ac.in}
  \and
  
 \IEEEauthorblockN{Aditya Siripuram}
  \IEEEauthorblockA{Email: staditya@ee.iith.ac.in}
  \and
  
 \IEEEauthorblockN{Brad Osgood}
  \IEEEauthorblockA{Email: osgood@stanford.edu}
}

\maketitle


\begin{abstract}
We consider the problem of building numerically stable algorithms for computing Discrete Fourier Transform (DFT) of $N$- length signals with known frequency support of size $k$. A typical algorithm, in this case, would involve solving (possibly poorly conditioned) system of equations, causing numerical instability. When $N$ is a power of 2, and the frequency support is a random subset of $\Z_N$, we provide an algorithm that has (a possibly optimal) $O(k \log k)$ complexity to compute the DFT, while solving system of equations that are $O(1)$ in size.
\end{abstract}

\section{Introduction}
This work deals with the computation of the Discrete Fourier Transform of signals whose frequency domain support is known beforehand. 
Let $N$ be a power of $2$, $\Z_N$ denote the ring of integers modulo $N$, and $\F:\mathbb{C}^N\mapsto \mathbb{C}^N$ denote the $N-$point Discrete Fourier transform (DFT) defined as
\[
\F f(m) = \sum_{n\in \Z_N}f(n)e^{-2\pi i m n /N}, \text{ for }m\in\Z_N.
\] 
We often refer to $f$ as the time-domain signal. Suppose $\JJ\subseteq \Z_N$ denotes a set of frequencies, and $\BB^\JJ$ denotes the subspace of signals $f$ whose DFT $\F f$ is zero outside $\JJ$. 
\begin{quote}
   (DFT computation with known support) Given sample access to $f \in \BB^\JJ$, and the frequency support $\JJ$, how do we compute the DFT coefficients $\F f$?
\end{quote}
 We denote by $k$ the size of the frequency support ($k=|\JJ|$). We can readily give the following straightforward solution to this problem. From the definition of the DFT, we set\footnote{For $\JJ \subseteq \Z_N$ and $f \in \mathbb{C}^N$, we represent by $f_\JJ$ the vector obtained by keeping only the elements indexed $\JJ$ from $f$. Similarly if $A$ is an $N\times N$ matrix whose rows/columns are indexed by elements of $\Z_N$, and $\II\in \mathbb{Z}_N$, we denote by $A(\II, \JJ)$ the submatrix of $A$ formed with columns $\JJ$ and rows $\II$. }
\(
f=\F^{-1}(\Z_N,\JJ)(\F f)_\JJ;
\)
from here, taking samples of the signal $f$ at $\II \subseteq \Z_N$ gives
\begin{equation}
f_\II = \F^{-1}(\II,\JJ)(\F f)_\JJ.
\label{eq:SoE}
\end{equation}
This gives a system of equations in the unknowns $(\F f)_\JJ$. Setting $\II = \{0,1,2,\ldots,k-1\}$ ensures the submatrix $\F^{-1}(\II,\JJ)$ is Vandermonde and invertible \cite{donoho-stark:uncertainty, DS-1}, and so this system can be solved in $O(k^2)$ arithmetic operations \cite{parker1964inverses,gohberg1997fast}. 

However, with some structural assumptions on the support $\JJ$, it might be possible to improve on this significantly \cite{10308632}. For example, when the DFT support is $\JJ = \{0,1,2,\ldots,k-1\}$ pick the time domain samples $\II$ to be uniformly spaced (i.e. \(\II=\{0,(N/k),2(N/k),\ldots\}\)), then the inverse of the submatrix $\F^{-1}(\II,\JJ)$ in equation \eqref{eq:SoE} is $k \times k$ inverse DFT matrix. Hence, solving \eqref{eq:SoE} is equivalent to computing the DFT of $f_\II$, and so can be done in $O(k \log k)$ arithmetic operations using Fast Fourier Transform (FFT) \cite{osgood2018lectures,cooley1965algorithm, good1958interaction, rader1968discrete}. Indeed, when the frequency support is the entire set of frequencies (i.e., $k=N$), the best-known complexity is $O(N\log N)$.

We draw the following impressions from the discussion above. First, depending on the structure of $\JJ$, we may wish to target an $O(k\log k)$ (rather than $O(k^2)$) algorithm to compute the DFT coefficients. Secondly, we may hope to investigate algorithms that potentially combine the system of equations approach (as in \eqref{eq:SoE}: this approach works for arbitrary frequency support structures $\JJ$) and the smaller FFT computation approach (that work for specific, structured, $\JJ$, as in the example above). This work, similar to our previous work \cite{10308632}, deals with using system of equation solvers and FFT sub-blocks to build a solution to the aforementioned problem of DFT computation with known frequency support.

\begin{figure}[ht]
\centering
\includegraphics[height=0.17\textwidth,width=0.32\textwidth]{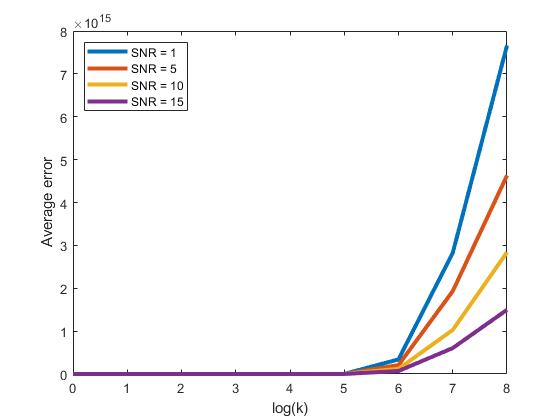}
\caption{\scriptsize The Average error of the submatrix method concerning size (log scale) of the submatrix $\log k$. Here, the length of the discrete signal is $N = 2^{14}$. Each point on the plot is obtained by averaging over 10000 runs. This is with the standard $O(k^3)$ matrix inversion. Using the $O(k^2)$ technique from \cite{parker1964inverses} seems to be worse regarding average error.}
\label{fig:submatirx}
\end{figure}

In contrast with the prior work \cite{10308632}, this work aims to build numerically stable algorithms for DFT computation with known support. For example, the $O(k^2)$ solution discussed in \eqref{eq:SoE}: this method involves solving a $k\times k$ Vandermonde system. This could have numerical stability issues due to poor conditioning of these submatrices: indeed, it has been observed that the conditioning of square Fourier submatrices scales exponentially with the size of the submatrix \cite{Pan2016}. We can see this in Fig. \ref{fig:submatirx}: this plots the average error (average $l_2$ norm of the difference between the original and estimated DFT coefficients over 10000 runs) with the size of the submatrix\footnote{In each run at a given $k$, set $\JJ$ is generated to be a random subset of $\Z_N$ of size $k$, and then the estimates are computed by solving equation (\ref{eq:SoE}) using standard $O(k^3)$ inversion.
}. 

We surmise that the size of these submatrices/system of equations involved in a DFT algorithm plays an important role in the numerical stability of the computations. The goal of this work is to investigate solutions to the DFT computation problem stated above by 1) combining system of equation solvers and FFT subblocks, 2) keeping the sizes of the involved system of equations as small as possible, and 3) achieving a target complexity of $O(k\log k)$.

We discuss the problem statement in more detail in Section \ref{sec:problem-set-up} and compare it with related work in Section \ref{sec:related-work}. We then discuss some suboptimal solutions based on our previous work in Section \ref{sec:shift-and-sample}. In Section \ref{sec:shift-and-progressive}, we propose a new algorithm (when the support sets $\JJ$ are random subsets of $\Z_N$) that achieves the target $O(k\log k)$ complexity by using system of equation solvers of size $O(1)$.

\section{Problem setup and result summary}
\label{sec:problem-set-up}
Given sample access to $f \in \BB^\JJ$, and $\JJ$, a \emph{structured DFT (SDFT) algorithm} for $\BB^\JJ$ outputs the DFT coefficients $(\F f)_\JJ$.


As motivated in the introduction, we refer to a SoE-SDFT algorithm as a structured DFT algorithm that uses the following subblocks (along with addition and subtraction): 1) $m-$ point FFTs, and 2) Solving system of equations. An SoE-SDFT algorithm can have multiple such sub-blocks. We evaluate these algorithms on the following key metrics
\begin{itemize}
    \item Computational complexity of the algorithm is the total number of arithmetic operations (complex additions, multiplications, etc) used by the algorithm; and
    \item Numerical stability of an SoE-SDFT algorithm is taken as the average size of the system of equation sub-blocks involved in the algorithm. 
  \end{itemize}
  As discussed in the introduction, for Fourier submatrices, condition number generally scales poorly with the submatrix size \cite{Pan2016}. This motivates the definition of numerical stability above.
\begin{defn}
    We call an SoE-SDFT algorithm an $(R, C)$ SoE-SDFT algorithm for $\BB^\JJ$ if, for all inputs $f\in \BB^\JJ$, 1) the algorithm uses $O(C)$ arithmetic operations, and 2) the average size of the system of equations subblocks in the algorithm is $O(R)$.
\end{defn}
For any $\JJ$, by picking $\II= \{0,1,2,\ldots,k-1\}$ we obtain a $(k,k^2)$ SoE-SDFT algorithm by solving \eqref{eq:SoE}. The goal is to explore algorithms under some structural conditions on $\JJ$; which ideally result in $R=1$ and $C=k\log k$.

As an initial problem, we consider the case when the frequency support $\JJ$ is a random subset of $\Z_N$: each element of $\Z_N$ is included in the subset $\JJ$ with probability $k/N$, independent of the choice for all the other elements. We refer to this distribution on $\JJ$ as $\Z_N^{\downarrow k}$. 

The key contribution of this work is an $(1,k\log k)$ SoE-SDFT algorithm for $\BB^\JJ$ when $\JJ \sim \Z_N^{\downarrow k}$. This builds on the $(\log k, k \log k)$ and $(1, k\log^2 k)$ algorithms that result as a byproduct of our earlier work \cite{10308632}. However, note that the earlier work in \cite{10308632} did not focus on numerical stability or matrix sizes; the only relevant metric for \cite{10308632} is computational complexity.

Though we start with random subsets of $\Z_N$, we hope to be able to extend these results to the case when $\JJ$ has a more general additive structure, similar to our prior work \cite{10308632}.  

\section{Related work}
\label{sec:related-work}
 The importance of the aforementioned problem comes from its applications in spectrum sensing \cite{3055462}, Sparse FFT \cite{gilbert2014recent} (which can be used as a subroutine to identify the DFT coefficient values) and its connection to the Fourier set query problem \cite{Kap17, GSS22, SSWZ22}. 
 
To our knowledge, this problem has not been investigated in detail in the literature. Following up on the simple linear algebraic solution discussed \eqref{eq:SoE}, we may consider using more time domain samples \cite{candes:robust} to obtain a near unitary tall matrix to improve the numerical stability of \eqref{eq:SoE}. But the computational complexity in this case is $\Omega(k^2)$, exceeding our target of $\tilde{O}(k)$.

A closely connected problem well studied in the literature is Sparse FFT (SFFT), where the support is unknown. We can also apply the Sparse FFT (SFFT) algorithms to the problem under investigation (by ignoring the available support information) to get an $O(k \log k)$ complexity \cite{pawar-ramachandran} \cite{ghazi2013sample}. However, these algorithms rely heavily on the Chinese Remainder theorem for isolating frequency coefficients. They do not work for the case when $N$ is a power of $2$ as in the proposed work. We can consider resampling the signal to $\hat{N}$ samples, where $\hat{N}$ has multiple prime factors, and then apply these algorithms. However, this resampling may incur an additional approximation and computation cost. Also, to the best of our knowledge, the conditioning of these methods has not been investigated. We hope the present technique generalizes to other support structures (not just random subsets of $\Z_N$), including additive structures \cite{10308632}.
\section{Congruence trees}
\label{sec:trees}
\begin{figure}
\centering
\resizebox{0.45\columnwidth}{!}{%
\begin{tikzpicture}[scale=0.5]
\node (0) at (0,8) [ minimum size = 0.25cm, fill=gray!10, draw] {$\scriptscriptstyle{\substack{\{0,1,6,7,38,65,135,512\}}}$};
\node (1.1) at (-4,5) [ minimum size = 0.25cm, fill=gray!10, draw] {$\scriptscriptstyle{\substack{\{1,7,64,135\}}}$};
\node (1.2) at (4,5) [ minimum size = 0.25cm, fill=gray!10, draw] {$\scriptscriptstyle{\substack{\{0,6,38,512\}}}$};
\node (2.1) at (-6,2) [ minimum size = 0.25cm, fill=gray!10, draw] {$\scriptscriptstyle{\substack{\{7,135\}}}$};
\node (2.2) at (-2,2) [ minimum size = 0.25cm, fill=gray!10, draw] {$\scriptscriptstyle{\substack{\{1,65\}}}$};
\node (2.3) at (2,2) [ minimum size = 0.25cm, fill=gray!10, draw] {$\scriptscriptstyle{\substack{\{6,38\}}}$};
\node (2.4) at (6,2) [ minimum size = 0.25cm, fill=gray!10, draw] {$\scriptscriptstyle{\substack{\{0,512\}}}$};

\draw (node cs:name=0) --node[above left] {$\scriptscriptstyle{\text{mod }2}$} (node cs:name =1.1);
\draw (node cs:name=0) -- node[above right] {$\scriptscriptstyle{\text{mod }2}$}(node cs:name =1.2);

\draw (node cs:name=1.1) --node[above left] {$\scriptscriptstyle{\text{mod }4}$} (node cs:name =2.1);
\draw (node cs:name=1.1) --node[above right] {$\scriptscriptstyle{\text{mod }4}$} (node cs:name =2.2);
\draw (node cs:name=1.2) --node[above left] {$\scriptscriptstyle{\text{mod }4}$} (node cs:name =2.3);
\draw (node cs:name=1.2) --node[above right] {$\scriptscriptstyle{\text{mod }4}$} (node cs:name =2.4);


\end{tikzpicture}}
\caption{\scriptsize Congruence tree for set $\{0,1,6,7,38,65,135,512\}$ obtained by removing all the nodes below level 2. Here $N=1024$ and $\mu_r^* = 2$ at level 2. The shift and sample algorithm uses two 4-point DFTs and four $2\times2$ systems to compute $\F f_{\JJ}$.}
\label{fig:shift-sample-example}
\end{figure}
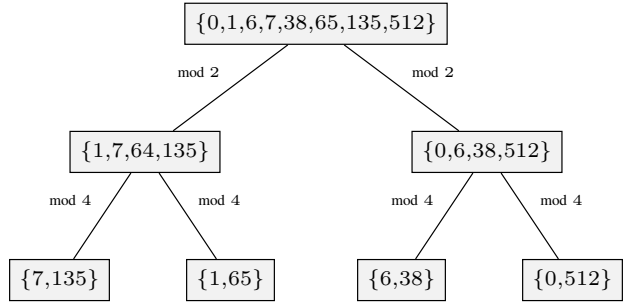

We set up the notation required to explain the algorithm. 
We use the following tree-based representation of support sets $\JJ \in \Z_N$ (similar to \cite{10308632} \cite{kapralov2019dimension}). The congruence tree of any $\JJ \subseteq \Z_N$ is a binary tree (with nodes labeled by the subsets of $\JJ$) obtained by splitting $\JJ$ into congruence classes modulo increasing powers of $2$. The congruence tree of $\JJ$ is defined recursively as  $(\text{left subtree}, \text{right subtree}, \text{root label})$:
 \[
\TT(\JJ) = \left(\TT(\JJ_{\text{odd}}) , \TT(\JJ_{\text{even}}), \JJ \right),
\]
where $\JJ_{\text{even}}, \JJ_{\text{odd}} \subseteq \Z_N$ describe the even and odd indices in $\JJ$ via  $2\JJ_{\text{even}}\subseteq \JJ$, $2\JJ_{\text{odd}}+1\subseteq \JJ$ respectively. At a level of $r$ from the root, the congruence tree of a set splits the set modulo $2^r$ (for example: see Fig. \ref{fig:shift-sample-example}). For any node $v$ at level $r$ in $\TT(\JJ)$, we denote by $\mu_\JJ(v)$ the size of the subset labeling node $v$. We will often need to work with the largest size label at level $r$, defined as $\mu_r^\star \coloneq \max_{v \text{ at level }r}\mu_\JJ(v)$. Lastly, the congruence tree for $\Z_N$ is denoted $\TT_N$. 

Consider the congruence tree $\TT(\JJ)$ and $\TT_N$. If $\JJ \sim \Z_N^{\downarrow k}$; then each leaf of $\TT_N$ is present in $\TT(\JJ)$ with probability $p = \frac{k}{N}$, and we have $2^{M-r}$ of such leaves contributing to each node at level $r$. Thus for any node $v$ at level $r$, $\mu_\JJ(v)
\sim \text{Binomial}(2^{M-r},p)$ is a Binomial random variable with mean $ \lambda \coloneqq 2^{M-r}p = k/2^{r}$. 

The probability that $\mu_\JJ(v)$ is large (or small) can be bounded by some well-known concentration inequalities \cite{mukhopadhyay2020probability}. For example, we can see that for $r = \lceil \log k \rceil$, $\mu_r^\star$ is $O(\log k)$ with high probability \cite{10308632}.

\section{Shift and sample}
\label{sec:shift-and-sample}
The following simple algorithm (investigated under a more general context in \cite{10308632}) can combine the two approaches referred to in the introduction. First, the congruence tree captures the aliasing pattern when an $f \in \BB^\JJ$ is sampled uniformly in the time domain. When $f \in \BB^\JJ$ is uniformly downsampled to $2^r$ samples ($f_{\downarrow 2^r}$), the DFT coefficients of the downsampled signal are given by 
\begin{equation}
\sum_{j\in \text{ label of }v} \F f(j), \text{ for each }v \text{ at level }r.
\label{eq:aliasing-eqns-node-wise}
\end{equation}
This can be seen by elementary techniques (e.g., \cite{osgood2018lectures}, \cite{10308632}). Thus, almost $\mu_r^\star$ DFT coefficients $\F f()$ overlap/add to result in the downsampled DFT. This motivates the shift and sample framework, explained with an example below. 
Consider the example set $\JJ=\{0,1,6,7,38,65,135,512\}$ and $N=1024$. Consider downsampling a signal $f \in \BB^\JJ$ to $4$ samples (so $r=2$ concerning the explanation above) and compute the $4-$point DFT of this downsampled signal. These computed values are equal to the sum of the corresponding overlapped coefficients at level $r=2$ in the congruence tree. For example, on the rightmost node, the frequency coefficients at indices $\{0,512\}$ overlap, and the computed $4-$point DFT value will be equal to $\F f(0) + \F f(512)$.

To resolve the values of $\F f(0)$ and $\F f(512)$ from the sum $\F f(0) + \F f(512)$, we can shift the time domain signal and repeat the above process to get new 4-point DFT values. This gives us the value of $\F f(0) + \F f (512) e^{512i/N}$. So we obtain a $2 \times 2$ Vandermonde system of equations in $\F f(0)$ and $\F f(512)$, which can be solved to obtain $\F f(0)$ and $\F f(512)$. Similarly, we can resolve the overlap at the three remaining nodes.

The idea of repeating the sampling process on shifts has been used repeatedly in DFT computation, see e.g., \cite{pawar-ramachandran,ghazi2013sample,nearlyoptimalsparse}.

The total number of computations, in this case, is $44$: the two $4-$point DFTs (one for the downsampled signal and one for the downsampled shifted signal) of size $4$ need $12$ computations each, and the four $2\times2$ system of equations can be solved each with $5$ computations. Compare this to the full DFT, which takes $15360$ arithmetic operations, or directly solving the system of equations, which needs $6 |\JJ|^2 = 384$ computations.

\usetikzlibrary{positioning,fit,calc} 
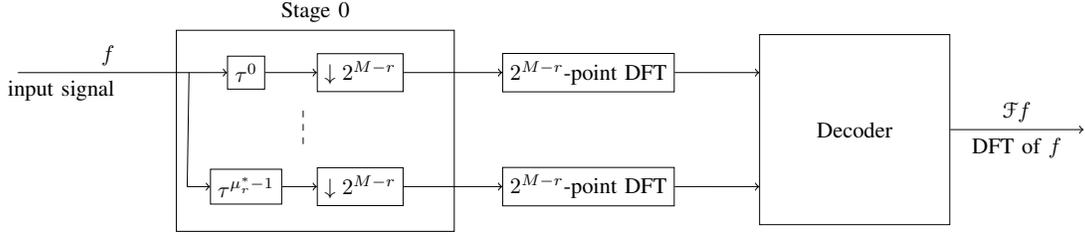
\begin{figure}[ht]
\centering
\resizebox{0.8\columnwidth}{!}{%
\begin{tikzpicture}[scale=0.6]

\node (s0_11) at (0,20) [ minimum size = 0.25cm, draw] {$\tau^0$};
\node (s0_12) at (3,20) [ minimum size = 0.25cm, draw] {$\downarrow 2^{M-r}$};
\node (s0_13) at (9,20) [ minimum size = 0.25cm, draw] {$2^{M-r}$-point DFT};
\draw[->] (node cs:name=s0_11) --node[above left] {} (node cs:name =s0_12); 
\draw[->] (node cs:name=s0_12) --node[above left] {} (node cs:name =s0_13);

\node (s0_21) at (0,17) [ minimum size = 0.25cm, draw] {$\tau^{\mu_r^*-1}$};
\node (s0_22) at (3,17) [ minimum size = 0.25cm, draw] {$\downarrow 2^{M-r}$};
\node (s0_23) at (9,17) [ minimum size = 0.25cm, draw] {$2^{M-r}$-point DFT};
\draw[->] (node cs:name=s0_21) --node[above left] {} (node cs:name =s0_22); 
\draw[->] (node cs:name=s0_22) --node[above left] {} (node cs:name =s0_23);

\draw [dashed](1.5,19) --node[above right] {} (1.5,18);
\node (s_0) [draw,inner xsep=8mm,inner ysep=4mm,fit=(s0_11) (s0_22),label={above:Stage 0}]{};

\draw[->] (-6,20) --node[above left] {$f$} node[below left] {input signal} (s0_11);
\draw[->] ($(s0_11)+(-1.5,0)$) -- ($(s0_21.west)+(-0.6,0)$) |- (s0_21.west);

\node (s_dec) at (16,18.5) [ minimum width=3cm,minimum height=3cm, draw] {Decoder};

\draw[->] (s0_13) -- ($(s_dec.west)+(0,1.5)$);
\draw[->] (s0_23) -- ($(s_dec.west)+(0,-1.5)$);

\draw[->] (s_dec.east) --node[above] {$\F f$} node[below] {DFT of $f$} ($(s_dec.east)+(3.5,0)$);
\end{tikzpicture}
}
\caption{\scriptsize Framework for Shift and Sample Algorithm. The signal is repeatedly shifted and downsampled ($\mu_r^*$ times). The decoder involves solving the Vandermonde system of equations involving the Fourier submatrix. We use $\tau$ to represent the shift operator ($\tau f (n) = f(n-1)$).
}
\label{fig:framework_alg1}
\end{figure}

Next, we summarize the relevant observations from \cite{10308632} resulting from the choice of the parameter $r$. Note that $r$ controls the tradeoff between the size of the DFT subblocks and the size of the system of equations subblocks. Setting $r=0$ reverts this algorithm to the submatrix method (equation (\ref{eq:SoE}) from the Introduction), and setting $r=\log M$ forces a full $N-$point DFT computation (and no system of equation sub-blocks).

By picking $r_* = \lceil \log k - \log \log k\rceil$, it can be shown that this shift and sample algorithm computes the DFT of signals in $\BB^\JJ$ (for $\JJ \sim \Z_N^{\downarrow k}$) using $O(k \log k)$ arithmetic operations, thus leading to a potentially optimal algorithm \cite{10308632}. However, this choice of $r$ does not necessarily lead to numerical stability in the sense we discussed earlier. The average sizes of the matrices involved are $E(\mu_\JJ(v)) = k/2^{r_*} = \log k$, and thus in the context of this work, this choice of $r$ gives an $(\log k, k \log k)$ SoE-SDFT algorithm. 

Similarly, picking $r^* = \lceil \log k\rceil$ results in a complexity of $O(k \log^2 k)$ (\cite[see eq (16)]{10308632}). However, in this scenario, the average sizes of the system of equations are $E(\mu_\JJ(v)) = k/2^{r^*} \leq 1$, thus giving a numerically stable algorithm. In the context of this work, this is a $(1, k \log^2k)$-SoE SDFT algorithm.

The present manuscript aims to present a $(1, k \log k)$ SoE-SDFT algorithm, thus combining the best attributes of the two algorithms above. Next, we explain the key idea behind the approach.

\section{Proposed algorithm: shift and progressive sample}
\label{sec:shift-and-progressive}In the first step of the shift and sample algorithm from Section \ref{sec:shift-and-sample}, we compute the DFT of the $f$ downsampled to $2^r$ points. We then repeated the process on the shifted signal $\tau f$ (by downsampling $\tau f$ to $2^r$ points). Instead, consider the following modification: each time we repeat the downsampling process to resolve overlaps, we increase the downsampling factor by $2$. For e.g. in the second step, we downsample $\tau f$ to only $2^{r-1}$ points (thus reducing the resolution in the time domain). The resulting DFT is twice as cheap to compute as the first step. Of course, this potentially leads to even more aliasing in the frequency domain due to the reduced sampling rate, thus increasing the size of the matrices involved (see Fig \ref{fig:tree_algo_2}).  This is the key idea behind the proposed Algorithm (Algorithm \ref{alg:shift-progressive-sample}).

\begin{algorithm}[ht]
\algsetup{linenosize=\tiny}
  \scriptsize
\caption{Shift and Progressive Sample Algorithm}
\label{alg:shift-progressive-sample}
  \begin{algorithmic}
  \STATE \textbf{Input: } $f$: Input time domain signal , $r$: Level at which DFT is computed , $\TT(\JJ)$: Binary tree of support set $\JJ$ 
  \STATE \textbf{Output: } $\F f(\JJ)$: The DFT coefficients of the signal $f$ on $\JJ$
  \STATE 1: \textbf{for} $l = r$ to $0$
  \STATE 2: \hspace{0.25cm} Compute $\neqlvl$ DFTs of the shifted and sampled signal: $\F (\tau^{(r-l)\neqlvl}f)_{\downarrow2^{M-l}}$ to $\F (\tau^{(r-l+1)\neqlvl-1}f)_{\downarrow2^{M-l}}$;
  \STATE 3: \hspace{0.25cm} \textbf{for} $i = 0$ to $2^{l-1}$
  \STATE 4: \hspace{0.5cm} Follow steps 1),2),3), and 4) in section \ref{sec:shift-and-progressive-details};
  \STATE 5: \hspace{0.5cm} 
  \textbf{If} (the system at $v_i$ is solvable)
  \STATE 6: \hspace{0.75cm} solve the system $\nodemat_\JJ(v_i)$ at node $v_i$, level $l$ and mark the node resolved;
  \STATE 7: \hspace{0.5cm} \textbf{End If}
  \STATE 8: \hspace{0.25cm} \textbf{end for}
  \STATE 9: \textbf{end for}
    \end{algorithmic}
\end{algorithm}

The key insight is that at the initial stage, some fraction of the DFT coefficients will be unaliased and will be found by the algorithm. Thus, the number of unknown DFT coefficients has been reduced, and so in subsequent stages, the sampling rate may be reduced appropriately to match the reduced number of unknown DFT coefficients. In this broad sense, this idea is similar in spirit to the one used in \cite{nearlyoptimalsparse}. Our key result is that such progressive down-sampling to lower and lower resolutions does not cause any order increase in the size of the matrices while yielding a (possibly optimal) $O(k\log k)$ complexity.
\usetikzlibrary{positioning,fit,calc} 
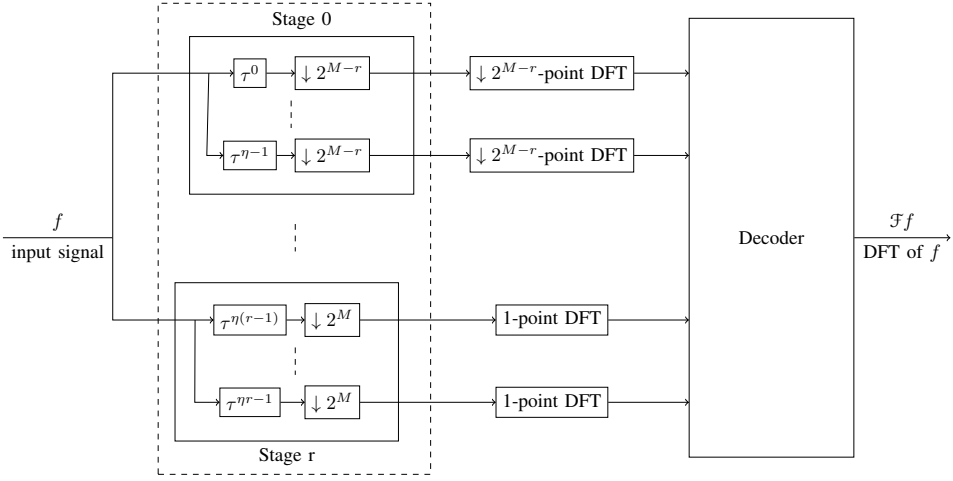
\begin{figure}[ht]
\centering
\resizebox{0.7\columnwidth}{!}{%
\begin{tikzpicture}[scale=0.5]

\node (s0_11) at (0,20) [ minimum size = 0.25cm, draw] {$\tau^0$};
\node (s0_12) at (3,20) [ minimum size = 0.25cm, draw] {$\downarrow 2^{M-r}$};
\node (s0_13) at (11,20) [ minimum size = 0.25cm, draw] {$\downarrow 2^{M-r}$-point DFT};
\draw[->] (node cs:name=s0_11) --node[above left] {} (node cs:name =s0_12); 
\draw[->] (node cs:name=s0_12) --node[above left] {} (node cs:name =s0_13);

\node (s0_21) at (0,17) [ minimum size = 0.25cm, draw] {$\tau^{\eta-1}$};
\node (s0_22) at (3,17) [ minimum size = 0.25cm, draw] {$\downarrow 2^{M-r}$};
\node (s0_23) at (11,17) [ minimum size = 0.25cm, draw] {$\downarrow 2^{M-r}$-point DFT};
\draw[->] (node cs:name=s0_21) --node[above left] {} (node cs:name =s0_22); 
\draw[->] (node cs:name=s0_22) --node[above left] {} (node cs:name =s0_23);

\draw [dashed](1.5,19) --node[above right] {} (1.5,18);
\node (s_0) [draw,inner xsep=8mm,inner ysep=4mm,fit=(s0_11) (s0_22),label={above:Stage 0}]{};


\node (sr_11) at (0,11) [ minimum size = 0.25cm, draw] {$\tau^{\eta (r-1)}$};
\node (sr_12) at (3,11) [ minimum size = 0.25cm, draw] {$\downarrow 2^M$};
\node (sr_13) at (11,11) [ minimum size = 0.25cm, draw] {1-point DFT};
\draw[->] (node cs:name=sr_11) --node[above left] {} (node cs:name =sr_12); 
\draw[->] (node cs:name=sr_12) --node[above left] {} (node cs:name =sr_13);

\node (sr_21) at (0,8) [ minimum size = 0.25cm, draw] {$\tau^{\eta r-1}$};
\node (sr_22) at (3,8) [ minimum size = 0.25cm, draw] {$\downarrow 2^M$};
\node (sr_23) at (11,8) [ minimum size = 0.25cm, draw] {1-point DFT};
\draw[->] (node cs:name=sr_21) --node[above left] {} (node cs:name =sr_22); 
\draw[->] (node cs:name=sr_22) --node[above left] {} (node cs:name =sr_23);

\draw [dashed](1.65,10) --node[above right] {} (1.65,9);
\node (s_r) [draw,inner xsep=7mm,inner ysep=4mm,fit=(sr_11) (sr_22),label={below:Stage r}]{};

\draw [dashed](1.65,14.5) --node[above right] {} (1.65,13.5);
\node (s_all) [dashed, draw,inner xsep=3mm,inner ysep=6mm,fit=(s_0) (s_r),label={}]{};

\draw (-9,14) --node[above] {$f$} node[below] {input signal} (-5,14);
\draw[->] (-5,14) -- ($(-5,14)$) |- (s0_11);
\draw[->] ($(s0_11)+(-1.5,0)$) -- ($(s0_21.west)+(-0.6,0)$) |- (s0_21.west);

\draw[->] (-5,14) -- ($(-5,14)$) |- (sr_11);
\draw[->] ($(sr_11)+(-2,0)$) -- ($(sr_21.west)+(-0.9,0)$) |- (sr_21.west);

\node (s_dec) at (19,14) [ minimum width=3cm,minimum height=8cm, draw] {Decoder};

\draw[->] (s0_13) -- ($(s_dec.west)+(0,6)$);
\draw[->] (s0_23) -- ($(s_dec.west)+(0,3)$);

\draw[->] (sr_13) -- ($(s_dec.west)+(0,-3)$);
\draw[->] (sr_23) -- ($(s_dec.west)+(0,-6)$);

\draw[->] (s_dec.east) --node[above] {$\F f$} node[below] {DFT of $f$} ($(s_dec.east)+(3.5,0)$);
\end{tikzpicture}
}
\caption{\scriptsize Framework for algorithm \ref{alg:shift-progressive-sample}. The input signal goes through multiple stages: in each stage, the signal is repeatedly shifted and downsampled (we shift $\eta$ times each stage, where $\eta$ is a constant integer). At stage $s$ (after some shift) the signal is downsampled uniformly by a factor of $2^{M-s}$. The decoder involves solving system of equations involving Fourier submatrix blocks. The sampling at the final stage takes just one sample.}
\label{fig:framework_alg2}
\end{figure}
\subsection{Details}
\label{sec:shift-and-progressive-details}
Suppose $j_1, j_2, \ldots, j_k$ are the elements in $\JJ$, recall the goal of the proposed algorithm is to find $\F f(j_1), \F f(j_2), \ldots, $. Initially, all these values are marked as unknown. 

Algorithm \ref{alg:shift-progressive-sample} operates in stages. At the initial (zeroth) stage, we operate at level $r = \left\lceil\log k\right\rceil$ of $\TT(\JJ)$, and in each subsequent stage, we move one level up (towards the root) of the tree. The algorithm visits all the tree nodes from $r$ upwards in order (thus, the algorithm has a total of $r$ stages).  As the algorithm unfolds, each node is linked to a system of equations (with variables as the unknowns corresponding to the node).

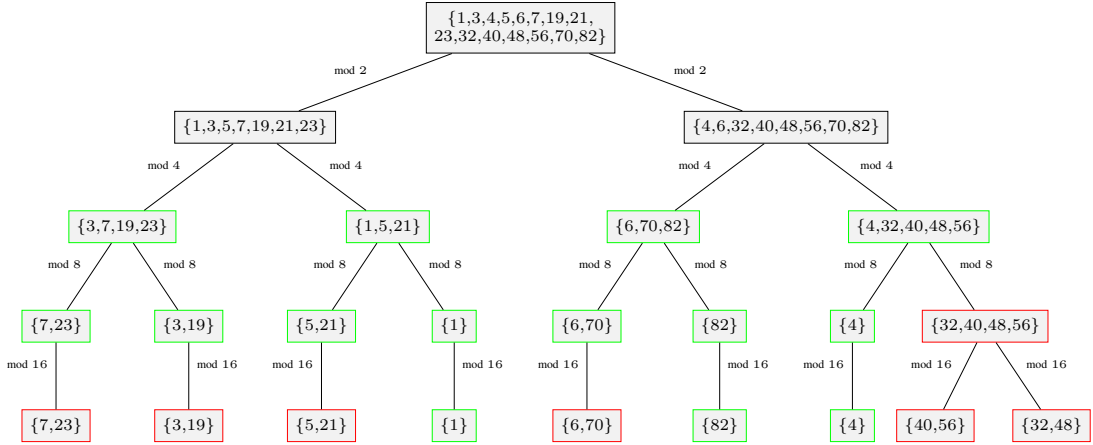
\begin{figure}[ht]
\centering
\resizebox{0.8\columnwidth}{!}{%
\begin{tikzpicture}[scale=0.5]
\node (0) at (0,8) [ minimum size = 0.25cm, fill=gray!10, draw] {$\scriptscriptstyle{\substack{\{1,3,4,5,6,7,19,21,\\23,32,40,48,56,70,82\}}}$};
\node (1.1) at (-8,5) [ minimum size = 0.25cm, fill=gray!10, draw] {$\scriptscriptstyle{\substack{\{1,3,5,7,19,21,23\}}}$};
\node (1.2) at (8,5) [ minimum size = 0.25cm, fill=gray!10, draw] {$\scriptscriptstyle{\substack{\{4,6,32,40,48,56,70,82\}}}$};
\node (2.1) at (-12,2) [ minimum size = 0.25cm, fill=gray!10, draw=green!100] {$\scriptscriptstyle{\substack{\{3,7,19,23\}}}$};
\node (2.2) at (-4,2) [ minimum size = 0.25cm, fill=gray!10, draw=green!100] {$\scriptscriptstyle{\substack{\{1,5,21\}}}$};
\node (2.3) at (4,2) [ minimum size = 0.25cm, fill=gray!10, draw=green!100] {$\scriptscriptstyle{\substack{\{6,70,82\}}}$};
\node (2.4) at (12,2) [ minimum size = 0.25cm, fill=gray!10, draw=green!100] {$\scriptscriptstyle{\substack{\{4,32,40,48,56\}}}$};
\node (3.1) at (-14,-1) [ minimum size = 0.25cm, fill=gray!10, draw=green!100] {$\scriptscriptstyle{\substack{\{7,23\}}}$};
\node (3.2) at (-10,-1) [ minimum size = 0.25cm, fill=gray!10, draw=green!100] {$\scriptscriptstyle{\substack{\{3,19\}}}$};
\node (3.3) at (-6,-1) [ minimum size = 0.25cm, fill=gray!10, draw=green!100] {$\scriptscriptstyle{\substack{\{5,21\}}}$};
\node (3.4) at (-2,-1) [ minimum size = 0.25cm, fill=gray!10, draw=green!100] {$\scriptscriptstyle{\substack{\{1\}}}$};
\node (3.5) at (2,-1) [ minimum size = 0.25cm, fill=gray!10, draw=green!100] {$\scriptscriptstyle{\substack{\{6,70\}}}$};
\node (3.6) at (6,-1) [ minimum size = 0.25cm, fill=gray!10, draw=green!100] {$\scriptscriptstyle{\substack{\{82\}}}$};
\node (3.7) at (10,-1) [ minimum size = 0.25cm, fill=gray!10, draw=green!100] {$\scriptscriptstyle{\substack{\{4\}}}$};
\node (3.8) at (14,-1) [ minimum size = 0.25cm, fill=gray!10, draw=red!100] {$\scriptscriptstyle{\substack{\{32,40,48,56\}}}$};
\node (4.1) at (-14,-4) [ minimum size = 0.25cm, fill=gray!10, draw=red!100] {$\scriptscriptstyle{\substack{\{7,23\}}}$};
\node (4.2) at (-10,-4) [ minimum size = 0.25cm, fill=gray!10, draw=red!100] {$\scriptscriptstyle{\substack{\{3,19\}}}$};
\node (4.3) at (-6,-4) [ minimum size = 0.25cm, fill=gray!10, draw=red!100] {$\scriptscriptstyle{\substack{\{5,21\}}}$};
\node (4.4) at (-2,-4) [ minimum size = 0.25cm, fill=gray!10, draw=green!100] {$\scriptscriptstyle{\substack{\{1\}}}$};
\node (4.5) at (2,-4) [ minimum size = 0.25cm, fill=gray!10, draw=red!100] {$\scriptscriptstyle{\substack{\{6,70\}}}$};
\node (4.6) at (6,-4) [ minimum size = 0.25cm, fill=gray!10, draw=green!100] {$\scriptscriptstyle{\substack{\{82\}}}$};
\node (4.7) at (10,-4) [ minimum size = 0.25cm, fill=gray!10, draw=green!100] {$\scriptscriptstyle{\substack{\{4\}}}$};
\node (4.8) at (12.5,-4) [ minimum size = 0.25cm, fill=gray!10, draw=red!100] {$\scriptscriptstyle{\substack{\{40,56\}}}$};
\node (4.9) at (16,-4) [ minimum size = 0.25cm, fill=gray!10, draw=red!100] {$\scriptscriptstyle{\substack{\{32,48\}}}$};

\draw (node cs:name=0) --node[above left] {$\scriptscriptstyle{\text{mod }2}$} (node cs:name =1.1);
\draw (node cs:name=0) -- node[above right] {$\scriptscriptstyle{\text{mod }2}$}(node cs:name =1.2);

\draw (node cs:name=1.1) --node[above left] {$\scriptscriptstyle{\text{mod }4}$} (node cs:name =2.1);
\draw (node cs:name=1.1) --node[above right] {$\scriptscriptstyle{\text{mod }4}$} (node cs:name =2.2);
\draw (node cs:name=1.2) --node[above left] {$\scriptscriptstyle{\text{mod }4}$} (node cs:name =2.3);
\draw (node cs:name=1.2) --node[above right] {$\scriptscriptstyle{\text{mod }4}$} (node cs:name =2.4);

\draw (node cs:name=2.1) --node[above left] {$\scriptscriptstyle{\text{mod }8}$} (node cs:name =3.1);
\draw (node cs:name=2.1) --node[above right] {$\scriptscriptstyle{\text{mod }8}$} (node cs:name =3.2);
\draw (node cs:name=2.2) --node[above left] {$\scriptscriptstyle{\text{mod }8}$} (node cs:name =3.3);
\draw (node cs:name=2.2) --node[above right] {$\scriptscriptstyle{\text{mod }8}$} (node cs:name =3.4);
\draw (node cs:name=2.3) --node[above left] {$\scriptscriptstyle{\text{mod }8}$} (node cs:name =3.5);
\draw (node cs:name=2.3) --node[above right] {$\scriptscriptstyle{\text{mod }8}$} (node cs:name =3.6);
\draw (node cs:name=2.4) --node[above left] {$\scriptscriptstyle{\text{mod }8}$} (node cs:name =3.7);
\draw (node cs:name=2.4) --node[above right] {$\scriptscriptstyle{\text{mod }8}$} (node cs:name =3.8);

\draw (node cs:name=3.1) --node[above left] {$\scriptscriptstyle{\text{mod }16}$} (node cs:name =4.1);
\draw (node cs:name=3.2) --node[above right] {$\scriptscriptstyle{\text{mod }16}$} (node cs:name =4.2);
\draw (node cs:name=3.3) --node[above left] {$\scriptscriptstyle{\text{mod }16}$} (node cs:name =4.3);
\draw (node cs:name=3.4) --node[above right] {$\scriptscriptstyle{\text{mod }16}$} (node cs:name =4.4);
\draw (node cs:name=3.5) --node[above left] {$\scriptscriptstyle{\text{mod }16}$} (node cs:name =4.5);
\draw (node cs:name=3.6) --node[above right] {$\scriptscriptstyle{\text{mod }16}$} (node cs:name =4.6);
\draw (node cs:name=3.7) --node[above left] {$\scriptscriptstyle{\text{mod }16}$} (node cs:name =4.7);
\draw (node cs:name=3.8) --node[above left] {$\scriptscriptstyle{\text{mod }16}$} (node cs:name =4.8);
\draw (node cs:name=3.8) --node[above right] {$\scriptscriptstyle{\text{mod }16}$} (node cs:name =4.9);

\end{tikzpicture}}
\caption{\scriptsize Successful execution of algorithm \ref{alg:shift-progressive-sample} on set: $\JJ = \{1,3,4,5,6,7,19,21,23,32,40,48,56,70,82\}$, $N=1024$, and $\neqlvl = 1$.}
\label{fig:tree_algo_2}
\end{figure}

Let $\neqlvl$ (a constant) be the number of DFTs computed by the algorithm at each stage. All the nodes in the tree are initialized as \emph{unresolved}. When at level $r$ (or stage $0$), the signal $f$ is shifted in the range $\neqlvl(r-l): \neqlvl(r-l +1) -1$, uniformly downsampled to $2^r$ samples for each shift, and the corresponding $2^r-$point DFTs computed. From \eqref{eq:aliasing-eqns-node-wise}, this gives $\eta$ equations at each of the $2^r$ nodes in level $r$. Likewise, in subsequent stages, say at level $l<r$ (stage $r-l$), the idea is to shift the signal $f$ in the range $\nodeshft(l) = [0:\neqlvl-1]$, uniformly downsampled to $2^l$ samples for each shift.


Thus, each tree node $\TT(\JJ)$ corresponds to a system of equations. We denote by $\nodemat_\JJ(i)$ the matrix for the system of equations at node $i$ and let $\nodelab_\JJ(i)$ denote the unknowns corresponding to node $i$. We may omit the subscript $\JJ$ when it is apparent from the context. For each node $i$ at level $l$,
\begin{enumerate}
    \item if both children are \emph{resolved}, then the obtained equation is redundant. We set $\nodemat(i) = NULL$, $\nodelab(i) = NULL$
    \item In case both children are \emph{unresolved}, then we have a \emph{merger}. The system at node $i$ is obtained by combining the systems at node $i_1$ and $i_2$. After this, $\neqlvl$ number of equations are taken by shifting the signal in the range $\nodeshft(i)=\neqlvl(r-l): \neqlvl(r-l +1) -1, $. To make this precise, assume $i_1$ is the left child and $i_2$ the right. We then have 
    \[
    \nodemat(i) =\left(\begin{array}{cc}
    \nodemat(i_1) & 0\\ 0 & \nodemat(i_2)\\ \multicolumn{2}{c}{\mathcal{F}(\nodeshft(i),\nodelab(i))} \end{array} \right), \quad \nodelab(i) = [\nodelab(i_1), \nodelab(i_2)].
    \]

    Which creates a new system of equations at node $i$ (with more unknowns than either of the children).
    \item In case one child is \emph{resolved} and the other \emph{unresolved}, the DFT value obtained at this node is adjusted by subtracting the part corresponding to the known (or \emph{resolved}) DFT coefficients. The system of equations at node $v$ has the same number of unknowns as the \emph{unresolved} child: the system from the unresolved child is propagated with some additional equations. Say $i_1, i_2$ are the children of $i$, with $i_2$ unresolved, we have 
     \[
    \nodemat(i) = \left(\begin{array}{cc}\nodemat(i_2) & 0 \\ \multicolumn{2}{c}{\mathcal{F}(\nodeshft(i) : ,\nodelab(i_2))} \end{array} \right), \quad \nodelab(i) = \nodelab(i_2)
    \]
    \item If $i$ is at level $r$ (or the zeroth stage of the algorithm), we set
    \[
    \nodemat(i) =  \mathcal{F}\left([0:\eta-1] ,\nodelab(i)\right),\quad \nodelab(i)=\text{ label of node }i.
    \]
    If $\nodemat(i)$ is tall (has more rows than columns), we clip it to square by removing the rows from the end. 
\end{enumerate}
  In cases 2), 3), and 4) above, the algorithm checks if the number of equations matches the number of unknowns. If yes, the node is marked \emph{resolved}, else the system of equations are propagated to subsequent stages. Once a node is marked resolved, the algorithm attempts to solve the resulting square system of equations. The algorithm reports a \emph{failure} if the square system is noninvertible. The algorithm terminates if we reach level $0$ (root) or if all the unknowns are solved. A successful run of the algorithm is when the root is \emph{resolved} and there are no \emph{failures}. 

 An example algorithm execution is shown in Fig. \ref{fig:tree_algo_2} for $\eta = 1$. All the nodes marked in green get resolved, and the nodes marked in red get propagated to the next stage.




\section{Analysis}

 We do the analysis (probabilistic) of the algorithm in two parts: 1) Correctness and 2) Complexity. In the correctness part, we show that the algorithm can identify all the DFT coefficients with high probability, and in the second part, we show that the average complexity is $O(k \log k)$. 

 Recall that we start the algorithm at level $r=\lceil \log k \rceil$ from the root in the congruence tree $\TT$. For simplicity, for the rest of the discussion, we remove all the nodes below this level from $\TT$ (so a leaf in $\TT$ will now be at level $r$).

\subsection{Correctness - part 1}
For any node $i$, we define the skewness of $i$ (written $\nodeskew_\JJ(i) $ or $\nodeskew(i)$ when $\JJ$ is apparent from the context) to be the difference of the number of columns and rows in the node matrix $\nodemat(i)$.
    \[
    \nodeskew(i) = \text{ncol}(\nodemat(i)) - \text{nrow}(\nodemat(i)).
    \]

Note that node $i$ is marked resolved iff $\nodeskew(i) = 0$.

\begin{defn}
We call a subtree $T$ of $\TT(\JJ)$ a \emph{merging tree} if 
\begin{enumerate}
    \item all nodes in $T$ (except possibly the root $T$) are unresolved (red), and
    \item For any unresolved (red) node in $T$, its unresolved (red) children are also in $T$.
\end{enumerate}

 The leaves of the merging tree are the unresolved nodes at the initial level (i.e., $\lceil \log k \rceil$) of the algorithm that constitutes the merging tree. The height of a merging tree is the height \footnote{The height of any node at level $l\leq r$ in a congruence tree is $r-l$. This is the number of levels above $r=\lceil \log k \rceil$ that the node is located.} of its root. We say that a merging tree is resolved if its root is resolved.
 \end{defn}
\begin{figure}[ht]
\centering
\resizebox{0.4\columnwidth}{!}{%
\begin{tikzpicture}[scale=0.5]
\node (0) at (0,6) [ minimum size = 0.25cm, fill=gray!10, draw=green!100] {$\scriptscriptstyle{\substack{\{32,40,48,56\}}}$};
\node (1) at (0,3) [ minimum size = 0.25cm, fill=gray!10, draw=red!100] {$\scriptscriptstyle{\substack{\{32,40,48,56\}}}$};
\node (2.1) at (-2,0) [ minimum size = 0.25cm, fill=gray!10, draw=red!100] {$\scriptscriptstyle{\substack{\{40,56\}}}$};
\node (2.2) at (2,0) [ minimum size = 0.25cm, fill=gray!10, draw=red!100] {$\scriptscriptstyle{\substack{\{32,48\}}}$};

\draw (node cs:name=0) --node[above left] {} (node cs:name =1);
\draw (node cs:name=1) --node[above left] {} (node cs:name =2.1);
\draw (node cs:name=1) --node[above right] {} (node cs:name =2.2);

\draw [|-|](5,6.5) --node[above right] {$\text{height}=2$} (5,-0.5);
\end{tikzpicture}}
\caption{\scriptsize Example merging tree from Fig.3 in the main draft.}
\label{fig:merging_tree_algo_2}
\end{figure}
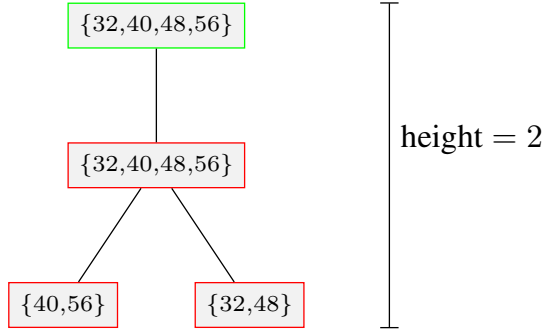

For example, for the set in Fig. \ref{fig:tree_algo_2}, the subtree on the right with leaves $\{40,56 \}$ and $\{32,48 \}$ is a merging tree with root $\{32,40,48,56\}$ as shown in Fig. \ref{fig:merging_tree_algo_2}. This tree has $2$ leaves and a height of $2$.

\begin{defn}
For any merging tree $T$, \begin{enumerate}
    \item the merging matrix of $T$, written $\nodemat(T)$ is defined as $\nodemat(T) \coloneqq \nodemat(\text{root(T)})$.

    \item The skewness of merging tree $T$, written $\nodeskew(T)$ is defined as $\nodeskew(T) \coloneqq \nodeskew(\text{root(T)})$. The skewness of $T$ represents how close $T$ is to being resolved.
    \item The weight of the merging tree $T$ is the number of leaves in the tree: $w(T)\coloneqq  \sum_{l \text{ leaf of } T} \mu_\JJ(l)$.
        \item We say that a merging tree $T$ is a complete (or resolved) merging tree if $\nodemat(T)$ is square, or equivalently, the root is resolved, or $\nodeskew(T)=0$. 
\end{enumerate} 

\end{defn}

We make the following simple observations
\begin{lemma}
\label{lem:merg_tree_properties}
Suppose $T$ is a merging tree in $\TT(\JJ)$.
\begin{enumerate}
    \item The columns in $\nodemat(T)$ correspond to the leaf node labels in order from the leftmost leaf to the right. The number of columns in $\nodemat(T)$ is $w(T)$ (=$\sum_{l \text{ leaf of } T} \mu_\JJ(l)$).
    
    \item The rows in $\nodemat(T)$ correspond to the equations taken at each stage of the algorithm. Generally, we expect $\eta$ rows corresponding to each node in $T$. The rows are listed according to the post-order traversal of the nodes in $T$. The matrix $\nodemat(T)$ is either wide (more columns than rows) or square. The number of rows in $\nodemat(T)$ is the smaller of $\neqlvl \times \text{ number of nodes in }T$ and $w(T)$.

\item For any node $i$, either the node is resolved (in which case $\nodeskew(i)=0$) or it is unresolved, in which case consider $T$ to be the merging tree rooted at $i$. Then $\nodeskew(i) = \nodeskew(T) = w(T) - \neqlvl \times \text{ number of nodes in }T$. 
\item For any node $i$, $\nodeskew(i)$ depends only on the weights of the descendants of $i$ in $\TT(\JJ)$. In particular, changing the weights of non-descendants (by adding or removing elements from $\JJ$) does not change the skewness $\nodeskew(i)$.
\item For any node $i$ with children $i_1,i_2$, we have $\nodeskew(i) = \max\{\nodeskew(i_1) + \nodeskew(i_2) - \eta, 0 \}$. If $i$ is a leaf node, then $\nodeskew(i) = \max\{ \mu(i) - \eta,0\}$.

\item If $T$ is a complete merging tree, then $1\leq \nodeskew(a) \leq \eta r$, and $\eta < \mu(a) \leq \eta (r+1)$ for any leaf $a$ of $T$.
   
\end{enumerate}
\end{lemma}

\begin{IEEEproof}
(1), (2), (3), and (4) follow directly from Definition 1 and Definition 2. (5) follow from (3) and (4) by noting the $\nodeskew$ values of the parent and the children.  

For (6), note that from Definition 2, in a complete merging tree, the skewness of any node except the root is non-zero. Hence the skewness of leaf $a$ is atleast 1 ($\nodeskew(a)\geq 1$). This implies $\mu(a)>\eta$. Starting from the root of the merging tree where $\nodeskew$ is 0, if we go to level one of the tree, the skewness will be $\eta$. Similarly, if we go to nodes at level $l$, the maximum skewness of a node at this level will be $\eta l$. The maximum height of a merging tree (distance from root to leaves) is $r$. And the maximum value of is ($\nodeskew(a)\leq \eta r$). This implies $\mu(a)\leq \eta(r+1)$.
\end{IEEEproof}

For the algorithm to find all the DFT coefficients, the root node of $\TT(\JJ)$  must be resolved. In particular, $\nodemat(\text{root})$ must be $NULL$ or square invertible. First, we show that the probability for $\nodemat(\text{root})$ to be wide (i.e., for $\nodemat(\text{root})$ to have more columns than rows) is small (Lemma \ref{lem:prob_height_unresolved} and Corollary \ref{cor:prob_unresolved_merg_tree}). This will establish that $\nodemat(\text{root})$ is either NULL (i.e., the root is resolved) or square, with high probability. Next, we show that any square $\nodemat(i)$ is invertible with high probability (Lemma \ref{lem:prob_inv_merg_matrix}). We will use these two to establish the correctness of the proposed algorithm (in Theorem \ref{thm:correctness_algo_2}).
\begin{lemma}
\label{lem:min_no_of_eqn}
Any merging tree with height $h$ and $y$ leaves has at least $2y-1+h-\log y$ nodes.
\end{lemma}
\begin{IEEEproof}
Suppose the number of nodes in the merging tree at level $i$ is given by $y_i$, for $i=0$ (leaf level) to $i=h$. We are given that $y_0=y$.  Since at most two nodes at a level can merge in the next (higher) level, we have $y_{i}/2 \leq y_{i+1}$. Also since $y_i \geq 1$, and   $y_{i+1} \leq y_i$, we have 
\[
\max\{1, y_i/2\} \leq y_{i+1} \leq y_i. 
\]
The total number of nodes is $\sum y_i$. It follows that the smallest number of nodes is achieved when $y_{i+1} = \lceil\max\{1, y_i/2\}\rceil$; leading to 
\begin{align*}
\text{ Min number of nodes } &\geq \underbrace{y + y/2 + y/4 + \ldots + 1 + 1 + 1}_{h \text{ times}}\\ 
&= y \left(2-\frac{1}{2^{\lfloor \log y \rfloor}}\right)+h-\lfloor \log y \rfloor-1\\
&\geq 2y-1+h-\log y
\end{align*}

\begin{figure}[ht]
\centering
\resizebox{0.6\columnwidth}{!}{%
\begin{tikzpicture}[scale=0.5]
\node (0) at (4,11) [ minimum size = 0.25cm, fill=gray!10, draw] {};
\node (1) at (4,8) [ minimum size = 0.25cm, fill=gray!10, draw] {};
\node (2) at (4,5) [ minimum size = 0.25cm, fill=gray!10, draw] {};
\node (3.1) at (0,2) [ minimum size = 0.25cm, fill=gray!10, draw] {};
\node (3.2) at (8,2) [ minimum size = 0.25cm, fill=gray!10, draw] {};
\node (4.1) at (-2,-1) [ minimum size = 0.25cm, fill=gray!10, draw] {};
\node (4.2) at (2,-1) [ minimum size = 0.25cm, fill=gray!10, draw] {};
\node (4.3) at (6,-1) [ minimum size = 0.25cm, fill=gray!10, draw] {};
\node (4.4) at (10,-1) [ minimum size = 0.25cm, fill=gray!10, draw] {};

\draw (node cs:name=0) --node[above left] {$\scriptscriptstyle{\text{mod }2}$} (node cs:name =1);

\draw (node cs:name=1) --node[above left] {$\scriptscriptstyle{\text{mod }4}$} (node cs:name =2);

\draw (node cs:name=2) --node[above left] {$\scriptscriptstyle{\text{mod }8}$} (node cs:name =3.1);
\draw (node cs:name=2) --node[above right] {$\scriptscriptstyle{\text{mod }8}$} (node cs:name =3.2);

\draw (node cs:name=3.1) --node[above left] {$\scriptscriptstyle{\text{mod }16}$} (node cs:name =4.1);
\draw (node cs:name=3.1) --node[above right] {$\scriptscriptstyle{\text{mod }16}$} (node cs:name =4.2);
\draw (node cs:name=3.2) --node[above left] {$\scriptscriptstyle{\text{mod }16}$} (node cs:name =4.3);
\draw (node cs:name=3.2) --node[above right] {$\scriptscriptstyle{\text{mod }16}$} (node cs:name =4.4);

\draw [|-|](-4,11) --node[above left] {$r = 4$} (-4,-1);
\draw [|-|](12,8) --node[above right] {$\text{height}=3$} (12,-1);

\end{tikzpicture}}

\caption{\scriptsize Example merging tree with $4$ leaves (with weight 2), here r = $4$ and merging tree height is $2$  .}
\label{fig:tree_alg_2}
\end{figure}
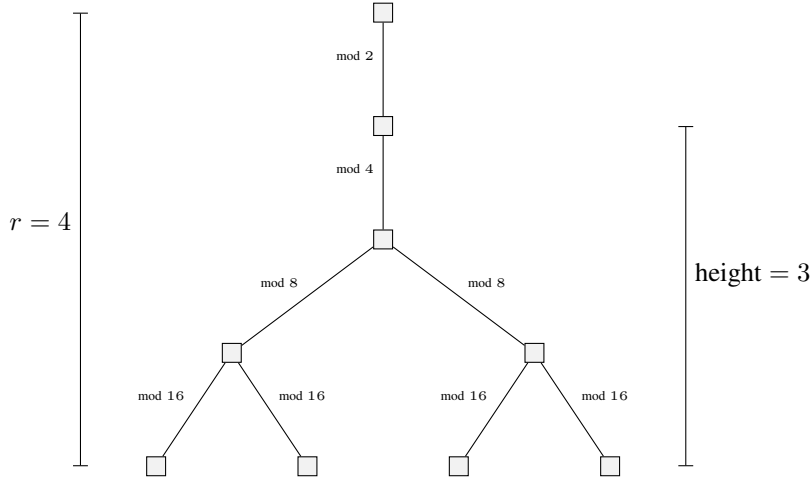

\end{IEEEproof}
Lemma \ref{lem:min_no_of_eqn} gives a lower bound the number of equations (i.e., rows in $\nodemat(T)$) obtained by the algorithm in a merging tree with a given number of leaves. Using this, we next show that $\nodemat(\text{root})$ cannot be wide with high probability.

\begin{lemma}
\label{lem:prob_height_unresolved}
For a node $v$ at a height $h$ in $\TT$,
\begin{enumerate}
    \item The probability that $v$ is unresolved decreases exponentially with the height of $v$: 
    \[
    \Prob( \nodeskew(v) \geq 1) \leq \exp{-\left(\frac{\neqlvl h-\lambda}{3}\right)}.
    \]
    \item The expected number of unknowns at $v$ decreases exponentially with the height of $v$     
    \[E(|\nodelab(v)|) \leq \exp\left( -\alpha_1 h \right), E(|\nodelab(v)|^2) \leq \exp\left( -\alpha_2 h \right), \text{ and }\quad E(|\nodelab(v)|^3) \leq \exp\left( -\alpha_3 h \right).
    \]
    \end{enumerate}

\end{lemma}
\begin{IEEEproof}
Suppose $v$ is unresolved, and let $T$ be the merging tree with $v$ as root. Let $i_1, i_2, \ldots i_y$ be the leaves of $T$.


Recall that $\mu_i \coloneqq \mu_\JJ(i)$ is the number of unknowns (amount of aliasing) at node  $i$. Now, for the first part of the proof, note that the number of unknowns in the system at the root is $\sum_{l=1}^{y} \mu_{i_l}$. Also note from Lemma \ref{lem:min_no_of_eqn} that we obtain at least $\neqlvl(2y-1+h-\log y)$ equations in these unknowns. Then, the probability that the system is underdetermined is upper bounded by
\[
P_{\textsf{under}}=\Prob\left( \sum_{l=1}^{y} \mu_{i_l} > \neqlvl(2y-1+h-\log y)\right).
\]

 We see that $\sum \mu_{i_l}$ is a sum of $y$, $N/2^r$iid Bernoulli random variables, each with mean $k/N$, so 
\[
E(\sum \mu_{i_l}) = y\lambda.
\]

we apply the concentration inequality from \eqref{eq:chernoff_upper} with $\delta = \neqlvl(2y-1+h-\log y)/y\lambda - 1$, to get

\begin{align*}
    P_{\textsf{under}} & \leq \exp{-\left(\frac{\neqlvl(2y-1+h-\log y) - y\lambda}{3}\right)} \\
    & \leq \exp{-\left(\frac{\neqlvl(y-1+h) - y\lambda}{3}\right)}  \text { using }\log y \leq y, \ y \geq 1\\
     &\leq \exp{-\left(\frac{\neqlvl h}{3}\right)}\exp{-\left(\frac{\eta(y-1)-y\lambda}{3}\right)}. 
\end{align*}

Next, we move to Lemma \ref{lem:prob_height_unresolved} Part 2. By the symmetry in our probability model, we expect that $E|(\nodelab(v)|^j)$ will be the same for all nodes $v$ at a given height; i.e., $E|(\nodelab(v)|^j)=u^j_h$ depends only on the height $h$ of $v$.

To start, note the following trivial bounds: the value of $|\nodelab(v)|$ cannot be larger than the total number of variables in the subtree rooted at $v$. Thus
\[
|\nodelab(v)| \leq \sum_{\substack{l \text{ is leaf of}\\v \text{ in }\TT}} I_l,
\]
where $I_l$ is indicator of whether an element $l \in \mathbb{Z}_N$ is included in $\JJ$ or not. 
The key intuition for the proof is that the value of $|\nodelab(v)|$ increases at least exponentially in $h$ (since the number of leaves of $v$ is exponential in $h$), but the probability of a node being unresolved itself decreases exponentially in $h$. Thus, with the correct rates, the value of $E(|\nodelab(v)|)$ will decrease exponentially in $h$.

Recall that $I_l$ are i.i.d with $E(I_l) = p=k/N$. A node $v$ has $N/2^{r-h}$ leaves, and so on the right-hand side in the above equation, $\sum I_l $ is a binomial random variable with parameters $N/2^{r-h}$ and $p$. The mean $M$ of this binomial random variable satisfies $2^{h-1} \leq M=N/2^{r-h} p = 2^hk/2^r \leq 2^h$. Then for $j=1,2,3,\ldots$, the $j^{th}$ moment $|\nodelab(v)|^j$ can be bounded by the $j^{th}$ moment of the binomial random variable on the right. In particular, we use the following bound on moments of a binomial random variable (see for e.g. \cite{berend2010improved,ahle2022sharp})
\[
E(X^j) \leq M^j \exp(j^2/2M), \text{ for binomial r.v. }X \text{ with mean } M.
\]
Using this (along with $2^{h-1}< M \leq  2^h$), we have
\begin{equation}
\label{eq:binom-moments-bound}
u^j_h=E|\nodelab(v)|^j \leq E\left(\sum_{\substack{l \text{ is leaf of}\\v \text{ in }\TT}} I_l\right)^j \leq 2^{hj} \exp(j^2/2^h) \leq e^{j^2} 2^{hj}.
\end{equation}

Let $I_v$ be the indicator variable which is $1$ if $v$ is unresolved. Consider the following recursive relationship for $|\nodelab(v)|$. For any non-leaf node $v$ of $\TT_r$ with children $v_1, v_2$,
\begin{equation}
\label{eq:node-unknown-recursion}
|\nodelab(v)| = |\nodelab(v_1)|I_{v_1} + |\nodelab(v_2)|I_{v_2}.
\end{equation}
First, note that by the Cauchy-Schwarz inequality for any node $v$ at height $h$,
\begin{align}
\label{eq:expec-cs-inequality}
E( |\nodelab(v)|I_{v} ) &\leq \left(E( |\nodelab(v)|^2 E(I_{v}) \right)^{1/2} \nonumber\\&= e^{-\frac{\eta h -1}{6}}  (u_h^2)^{1/2}
\end{align}
Taking expectation on \eqref{eq:node-unknown-recursion} and applying \eqref{eq:expec-cs-inequality},  we get
\begin{align*}
u_h &= E(|\nodelab(v)|) = E(|\nodelab(v_1)| I_{v_1}) + E(|\nodelab(v_2)|I_{v_2})  \\
&\leq (2u_{h-1}^2 E(I_{v_1}))^{1/2} \text{ from \eqref{eq:expec-cs-inequality}}\\
&\leq 2(u_{h-1}^2 e^{-\frac{\eta (h-1) -1}{3}})^{1/2}  \text{ from Lemma \ref{lem:prob_height_unresolved} part 1}\\
&\leq  2 e^{-\frac{\eta (h-1) -1}{6}} (u_{h-1}^2)^{1/2}\\
&\leq  2 e^{-\frac{\eta (h-1) -1}{6}} 2^{(h-1)}e^{\frac{2}{2^{h-1}}} \\
&\leq   e^2 2^{h} e^{-\frac{\eta (h-1) -1}{6}}   \\
&\leq e^(\frac{\eta+13}{6})  e^{-\frac{h(\eta-6 \ln 2)}{6}} 
\end{align*}
Taking the square of both sides in \eqref{eq:node-unknown-recursion},
\[
|\nodelab(v)|^2 = |\nodelab(v_1)|^2I_{v_1} + |\nodelab(v_2)|^2I_{v_2}+ 2|\nodelab(v_1)|I_{v_1}|\nodelab(v_2)|2I_{v_2}.
\]
Taking expectation and applying \eqref{eq:expec-cs-inequality},  we get
\begin{align*}
u_h^2 &= E|\nodelab(v)|^2 \\& = E(|\nodelab(v_1)|^2 I_{v_1}) + E(|\nodelab(v_2)|^2I_{v_2}) +  2 E(|\nodelab(v_1)|I_{v_1})E(|\nodelab(v_2)|I_{v_2}) \\
&\leq 2(u_{h-1}^4 E(I_{v_1}))^{1/2} +  2 E(|\nodelab(v_1)|I_{v_1})E(|\nodelab(v_2)|I_{v_2}) \\
&\leq 2(u_{h-1}^4 e^{-\frac{\eta (h-1) -1}{3}})^{1/2} + 2 e^{-\frac{\eta (h-1) -1}{3}}  u_{h-1}^2 \text{ from \eqref{eq:expec-cs-inequality}, Lemma \ref{lem:prob_height_unresolved} part 1}\\
&\leq  2 e^{-\frac{\eta (h-1) -1}{6}} ((u_{h-1}^4)^{1/2}+ e^{-\frac{\eta (h-1) -1}{6}}u_{h-1}^2) \\
&\leq  2 e^{-\frac{\eta (h-1) -1}{6}} (2^{(h-1)2}e^{8}+ e^{-\frac{\eta (h-1) -1}{6}}2^{(h-1)2}e^{4}) \\
&\leq  2^{2h-1} e^{-\frac{\eta (h-1) -1}{6}} e^4 (e^4+ e^{-\frac{\eta (h-1) -1}{6}}) \\
&\leq e^8 2^{2h}  e^{-\frac{\eta (h-1) -1}{6}} \\
&\leq e^(\frac{\eta+49}{6})  e^{-\frac{h(\eta-12 \ln 2)}{6}} 
\end{align*}

Taking the cube of both sides in \eqref{eq:node-unknown-recursion},
\[
|\nodelab(v)|^3 = |\nodelab(v_1)|^3I_{v_1} + |\nodelab(v_2)|^3I_{v_2} + 3|\nodelab(v_1)|^2I_{v_1}|\nodelab(v_2)|2I_{v_2}+3|\nodelab(v_1)|I_{v_1}|\nodelab(v_2)|^22I_{v_2}.
\]

Taking expectation and applying \eqref{eq:expec-cs-inequality},  we get
\begin{align*}
u_h^3 &= E|\nodelab(v)|^3\\
&= E(|\nodelab(v_1)|^3I_{v_1}) + E(|\nodelab(v_2)|^3I_{v_2}) + 3E(|\nodelab(v_1)|^2I_{v_1})E(|\nodelab(v_2)|2I_{v_2})+3E(|\nodelab(v_1)|I_{v_1})E(|\nodelab(v_2)|^22I_{v_2})\\
&\leq 2(u_{h-1}^6 E(I_{v_1}))^{1/2} + 6 (u_{h-1}^4 E(I_{v_1}))^{1/2}E(|\nodelab(v_2)|I_{v_2}) \text{ from \eqref{eq:expec-cs-inequality}}\\
&\leq 2 e^{-\frac{\eta (h-1) -1}{6}} (2^{3(h-1)}e^{\frac{18}{2^{h-1}}}+3 (2^{2(h-1)})e^{\frac{8}{2^{h-1}}}e^{-\frac{\eta (h-1) -1}{3}})\\
&\leq 2^{3h} e^{-\frac{\eta (h-1) -1}{6}} e^{18}\\
&\leq e^(\frac{\eta+109}{6})  e^{-\frac{h(\eta-18 \ln 2)}{6}}\text{ from Lemma \ref{lem:prob_height_unresolved} part 1}
\end{align*}




\end{IEEEproof}

\begin{corollary}
\label{cor:prob_unresolved_merg_tree}
The probability that the root of $\TT(\JJ)$ is unresolved when $\JJ \sim $ is upper bounded by $e^{\neqlvl/3}/k^{\neqlvl/3}$ 
\end{corollary}
\begin{IEEEproof}
This follows by substituting $r \geq \log k -1$ in Lemma \ref{lem:prob_height_unresolved}.
\end{IEEEproof}
Thus, with high probability, the algorithm obtains as many equations as unknowns before reaching the root. In the next section, we consider this square system of equations obtained by the algorithm and show that it is invertible with high probability.

\subsection{Correctness - part 2}

Consider a complete merging tree $T$ in $\TT(\JJ)$. Suppose $T$ has $y$ leaves, with labels  $\{a_{11}, a_{12}, \ldots\}$, $\{a_{21}, a_{22}, \ldots, \}$, $\ldots$, $\{a_{y1}, a_{y2}, \ldots \}$, with each $a_{ij} \in \JJ$. Suppose we remove one of the leaf elements from the leftmost leaf in $T$ (say $a_{11}$). Now we execute the algorithm on $\JJ'=\JJ \setminus \{a_{11}\}$. This leads to potentially new merging trees: in particular, the merging pattern from $T$ in $\TT(\JJ)$ may not sustain in $\TT(\JJ') $ because of one less unknown. Some of the nodes which were unresolved in $\TT(\JJ)$ may potentially be marked as resolved in $\TT(\JJ')$. 

\begin{lemma}
\label{lem:skewness_split}
Given a complete merging tree $T$ in $\TT(\JJ)$, let $\JJ'$ be obtained by removing an element from the leftmost leaf of $T$. Then, the skewness of any node $i$ is given by
\[
\nodeskew_{\JJ'}(i) = \begin{cases}\nodeskew_{\JJ}(i) \text{ if }i \text{ is not on the path from }a_{11}\text{ to root of }T \\\nodeskew_{\JJ'}(i) = \nodeskew_{\JJ}(i)-1 \text{ otherwise.}  \end{cases}
\]

\end{lemma}
\begin{IEEEproof}
\begin{figure}[ht]
\centering
\resizebox{0.7\columnwidth}{!}{%
\begin{tikzpicture}[scale=0.5]
\node[scale=0.5] (0.1) at (8,17) [ minimum size = 0.25cm, fill=gray!10, draw] {};
\node[scale=0.5] (0.2) at (8,15) [ minimum size = 0.25cm, fill=gray!10, draw] {};
\draw(node cs:name=0.1) --node[left] {$\scriptstyle{\nodeskew = 1}$} (node cs:name =0.2);

\node[scale=0.5] (0.3) at (8,13) [ minimum size = 0.25cm, fill=gray!10, draw] {};
\draw(node cs:name=0.2) --node[left] {$\scriptstyle{\nodeskew = 2}$} (node cs:name =0.3);

\node[scale=0.5] (0.4) at (8,11) [ minimum size = 0.25cm, fill=gray!10, draw] {};
\draw(node cs:name=0.3) --node[left] {$\scriptstyle{\nodeskew = 3}$} (node cs:name =0.4);

\node[scale=0.5] (1) at (6,9) [ minimum size = 0.25cm, fill=gray!10, draw] {};
\node[scale=0.5] (1.1) at (10,9) [ minimum size = 0.25cm, fill=gray!10, draw] {};

\draw(node cs:name=0.4) --node[above left] {$\scriptstyle{\nodeskew = 3}$} (node cs:name =1);

\draw(node cs:name=0.4) --node[above right] {$\scriptstyle{\nodeskew = 1}$} (node cs:name =1.1);

\node[scale=0.5] (2.1) at (4,7) [ minimum size = 0.25cm, fill=gray!10, draw] {};
\node[scale=0.5] (2.2) at (8,7) [ minimum size = 0.25cm, fill=gray!10, draw] {};

\draw (node cs:name=1) --node[above left] {$\scriptstyle{\nodeskew = 1}$} (node cs:name =2.1);
\draw (node cs:name=1) --node[above right] {$\scriptstyle{\nodeskew = 3}$} (node cs:name =2.2);

\node[scale=0.5] (3.1) at (2,5) [ minimum size = 0.25cm, fill=gray!10, draw] {};

\draw  (node cs:name=2.1) --node[above left] {$\scriptstyle{\nodeskew = 2}$} (node cs:name =3.1);

\node[scale=0.5] (4.1) at (0,3) [ minimum size = 0.25cm, fill=gray!10, draw] {};
\node[scale=0.5] (4.2) at (4,3) [ minimum size = 0.25cm, fill=gray!10, draw] {};

\draw (node cs:name=3.1) --node[above left] {$\scriptstyle{\nodeskew = 1}$} (node cs:name =4.1);
\draw (node cs:name=3.1) --node[above right] {$\scriptstyle{\nodeskew = 2}$} (node cs:name =4.2);
\node[scale=0.5] (5.1) at (-2,1) [ minimum size = 0.25cm, fill=gray!10, draw] {};

\draw (node cs:name=4.1) --node[above left] {$\scriptstyle{\nodeskew = 2}$} (node cs:name =5.1);

\node[scale=0.5] (6.1) at (-4,-1) [ minimum size = 0.25cm, fill=gray!10, draw] {};
\node[scale=0.5] (6.2) at (0,-1) [ minimum size = 0.25cm, fill=gray!10, draw] {};

\draw (node cs:name=5.1) --node[above left] {$\scriptstyle{\nodeskew = 1}$} (node cs:name =6.1);
\draw (node cs:name=5.1) --node[above right] {$\scriptstyle{\nodeskew = 2}$} (node cs:name =6.2);
\end{tikzpicture}
\hspace{2cm}
\begin{tikzpicture}[scale=0.5]
\node[scale=0.5] (0.1) at (8,17) [ minimum size = 0.25cm, fill=gray!10, draw] {};
\node[scale=0.5] (0.2) at (8,15) [ minimum size = 0.25cm, fill=gray!10, draw] {};
\draw[red,dashed] (node cs:name=0.1) --node[left] {\textcolor{green}{$\scriptstyle{\nodeskew = 0}$}} (node cs:name =0.2);

\node[scale=0.5] (0.3) at (8,13) [ minimum size = 0.25cm, fill=gray!10, draw] {};
\draw(node cs:name=0.2) --node[left] {$\scriptstyle{\nodeskew = 1}$} (node cs:name =0.3);

\node[scale=0.5] (0.4) at (8,11) [ minimum size = 0.25cm, fill=gray!10, draw] {};
\draw(node cs:name=0.3) --node[left] {$\scriptstyle{\nodeskew = 2}$} (node cs:name =0.4);

\node[scale=0.5] (1) at (6,9) [ minimum size = 0.25cm, fill=gray!10, draw] {};
\node[scale=0.5] (1.1) at (10,9) [ minimum size = 0.25cm, fill=gray!10, draw] {};

\draw(node cs:name=0.4) --node[above left] {$\scriptstyle{\nodeskew = 2}$} (node cs:name =1);

\draw(node cs:name=0.4) --node[above right] {$\scriptstyle{\nodeskew = 1}$} (node cs:name =1.1);

\node[scale=0.5] (2.1) at (4,7) [ minimum size = 0.25cm, fill=gray!10, draw] {};
\node[scale=0.5] (2.2) at (8,7) [ minimum size = 0.25cm, fill=gray!10, draw] {};

\draw[red,dashed] (node cs:name=1) --node[above left] {\textcolor{green}{$\scriptstyle{\nodeskew = 0}$}} (node cs:name =2.1);
\draw (node cs:name=1) --node[above right] {$\scriptstyle{\nodeskew = 3}$} (node cs:name =2.2);

\node[scale=0.5] (3.1) at (2,5) [ minimum size = 0.25cm, fill=gray!10, draw] {};

\draw  (node cs:name=2.1) --node[above left] {$\scriptstyle{\nodeskew = 1}$} (node cs:name =3.1);

\node[scale=0.5] (4.1) at (0,3) [ minimum size = 0.25cm, fill=gray!10, draw] {};
\node[scale=0.5] (4.2) at (4,3) [ minimum size = 0.25cm, fill=gray!10, draw] {};

\draw[red,dashed] (node cs:name=3.1) --node[above left] {\textcolor{green}{$\scriptstyle{\nodeskew = 0}$}} (node cs:name =4.1);
\draw (node cs:name=3.1) --node[above right] {$\scriptstyle{\nodeskew = 2}$} (node cs:name =4.2);
\node[scale=0.5] (5.1) at (-2,1) [ minimum size = 0.25cm, fill=gray!10, draw] {};

\draw (node cs:name=4.1) --node[above left] {$\scriptstyle{\nodeskew = 1}$} (node cs:name =5.1);

\node[scale=0.5] (6.1) at (-4,-1) [ minimum size = 0.25cm, fill=gray!10, draw] {};
\node[scale=0.5] (6.2) at (0,-1) [ minimum size = 0.25cm, fill=gray!10, draw] {};

\draw[red,dashed] (node cs:name=5.1) --node[above left] {\textcolor{green}{$\scriptstyle{\nodeskew = 0}$}} (node cs:name =6.1);
\draw (node cs:name=5.1) --node[above right] {$\scriptstyle{\nodeskew = 2}$} (node cs:name =6.2);
\end{tikzpicture}}

\caption{\scriptsize Example merging tree split}
\label{fig:s_m_tree}
\end{figure}
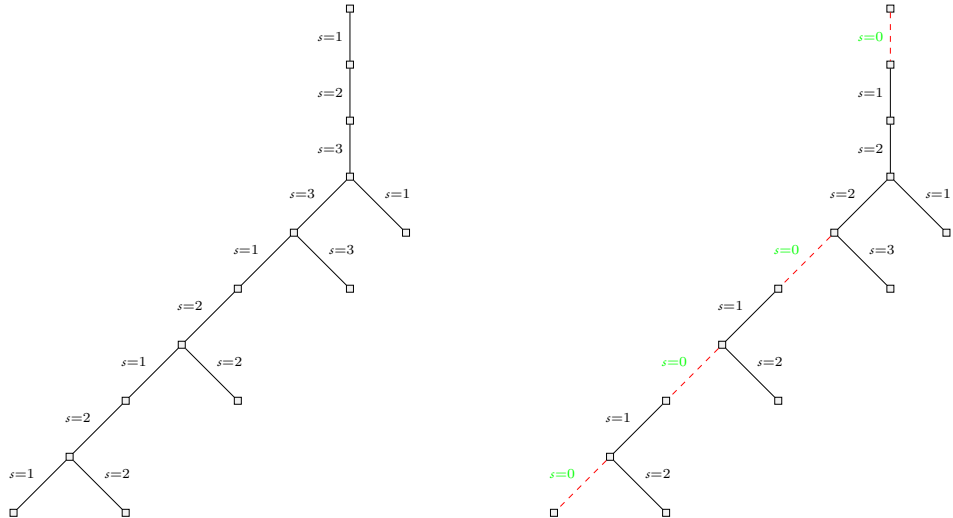
If $i$ is not on the path from $a_{11}$ to the root, this follows from Lemma \ref{lem:merg_tree_properties} (4). Otherwise, note that $\nodeskew(i) \geq 1$ by definition, and the desired relationship to $\nodeskew_\JJ'(i)$ follows by recursively applying Lemma \ref{lem:merg_tree_properties} (5) starting from the leaf node for $a_{11}$ and going towards the root of $T$. 
\end{IEEEproof}

We call any complete merging trees formed by any of the nodes $\{a_{12}, \ldots\}$, $\{a_{21}, a_{22}, \ldots, \}$, $\ldots$, $\{a_{y1}, a_{y2}, \ldots \}$ in $\TT(\JJ')$ as a split of $T$. We note the following properties as a corollary of Lemma \ref{lem:skewness_split}.

\begin{corollary}
\label{cor:split_tree_properties}
\begin{enumerate}
    \item Any node that is resolved in $\TT(\JJ)$ will be resolved in $\TT(\JJ')$ as well.
    \item Any split of $T$ is a subtree of $T$.
    \item Nodes not in the path from $a_{11}$ to the root that were unresolved in $\TT(\JJ)$ continue to be unresolved in $\TT(\JJ')$.
    \item Any split of $T$ with leaf node containing $a_{ij}$ will also include the common ancestor of $a_{ij}$ and $a_{11}$.
    \item The number of splits of $T$ is equal to the number of nodes $j$ in the path from $a_{11}$ to the root that satisfy $\nodeskew_\JJ(j) = 1$. In particular, any split of $T$ is rooted on the path from $a_{11}$ to $\text{root}(T)$, and all the splits are disjoint.
    \item All nodes in $T$, except possibly $\text{root}(T)$, are also in some split of $T$.
    \item Suppose $T$ has $s$ splits, $T_1, T_2, \ldots, T_s$ ordered according to how their roots are encountered on the path from $a_{11}$ to $\text{root}(T)$. We have \(\text{nodes(T)}= \begin{pmatrix}\text{nodes}(T_1) & \text{nodes}(T_2) & \ldots & \text{nodes}(T_s) & *\end{pmatrix} \), where $*$ is either root$(T)$ or empty (in which case $\text{root}(T) = \text{root}(T_s)$).
\end{enumerate}
\end{corollary}
\begin{IEEEproof}
(1). (3) and (4) follow directly from Lemma \ref{lem:skewness_split}. (2) follows from (1), for (5) note that the root of any split must be a node in $T$ and have zero skewness. From Lemma \ref{lem:skewness_split}, this is only possible for nodes in the path from $a_{11}$ to root that have skewness $1$. (6) follows from (5). For (7), note that in any post-order traversal of $T$, the nodes of $T_1$ are visited first, followed by the nodes of $T_2$, followed by $T_3$, and so on. Since we have removed one element from the leaves of $T$, the total weight of the splits must be one less than the total weight of $T$, i.e. $\sum w(T_i) = w(T)-1$, so that the number of nodes in $T_1, T_2, \ldots$ is at most one less than the number of nodes in $T$ (if $\eta=1$ then the number of nodes in $T_1, T_2, \ldots$ is exactly one less than the number of nodes in $T$).

\end{IEEEproof}

\begin{lemma}
\label{lem:structure_merg_matrix}
 Consider a complete merging tree $T$ in $\TT(\JJ)$. As discussed previously, let $\JJ'$ be obtained by removing an element from the leftmost leaf of $T$ from  $\JJ$. Suppose the tree $T$ splits into trees $T_1, T_2, \ldots$ in $\TT(\JJ')$, ordered according to Corollary \ref{cor:split_tree_properties}. Then the submatrix of $\nodemat_\JJ(T)$ is obtained by removing the first column, and the last row has the form \\ 
\resizebox{0.7\columnwidth}{!}{%
 $ \hspace{8cm}
\begin{pNiceArray}{ccccccccccccccc}[first-row,first-col]
  &  &   & & &   & & &   & & &   &  &  &  &   \\
  &  & \Block[draw,rounded-corners]{3-3}{}  &  &  & \Block{3-3}<\huge>{0}  & & &   & & &  \Block{3-3}<\huge>{0} &  &   &  &  \\
    &   &  & \huge{\nodemat_{\JJ'}(T_1)} &   & &  &   & &\dots  &   &  &   &  &  \\
  &  &   &  &  &   & & &   & & &   &  &   &  &  \\
   &  & \Block{3-3}<\huge>{*}  &  &  & \Block[draw,rounded-corners]{3-3}{}  & & &   && &  &  &   &  &  \\
    &   &  &  &   & & \huge{\nodemat_{\JJ'}(T_2)} &   & & &\ddots   &  & \vdots &  &   \\
  &  &   &  &  &   & & &   & & &  \Block{3-3}<\huge>{0} &  &   &  &  \\
  &  &    &  & \ddots &   & & &   & & &   &  &   &  &  \\
    &   &  & \vdots &   & &  &   & &\ddots &   &  &   &  &  \\
  &  & \Block{3-3}<\huge>{*}   &  &  &   & & & \Block{3-3}<\huge>{*}  & & &   & \Block[draw,rounded-corners]{3-3}{}  &   &  &  \\
  &   &  &  &   &  & \dots &   &  & &   &  &  \huge{\nodemat_{\JJ'}(T_s)} &  &  \\  
    &   &  &  &   & &  &   & & &   &  &   &  &    
\end{pNiceArray}
$}
\\
\end{lemma}

\begin{IEEEproof}
From Corollary \ref{cor:split_tree_properties} 6) above, note that in rows of the submatrix, the equations corresponding to nodes from $T_1$ appear first, followed by nodes from $T_2$ and so on. For any split $T_r$ we note that
\begin{enumerate}
    \item In the equations corresponding to $T_r$, the coefficients of the variables corresponding to leaves of $T_r$ are same for both $\JJ$ and $\JJ'$; thus the matrix in the diagonal blocks is $\nodemat_\JJ'(T_r)$.
    \item The equations corresponding to $T_r$ do not involve any variables from $T_s$ for $s>r$.
\end{enumerate}
This results in the structure above.
\end{IEEEproof}

Note that $\nodemat(T_i)$ are square matrices since $T_i$ are complete merging trees by definitions. Thus, the determinant of this submatrix is simply the product of the determinants of the diagonal blocks. For a given merging pattern, the matrix corresponding to this system of equations has the following form:

First, note that $\nodemat(T)$ is constructed from Fourier submatrices, and as such, all the entries in $\nodemat(T)$ are either zero or powers of $\zeta = \exp(2 \pi i /N)$. More specifically, all the entries in a particular column $j$ are either zero or powers of $\exp(2 \pi i a_j /N)$. Here $a_1,a_2,\ldots$ are the elements of leaf node labels. This leads naturally to the following definition.

\begin{defn}
For any complete merging tree, consider replacing $\exp(2 \pi i a_j /N)$ in column $j$ of $\nodemat(T)$,  with the variable $x_{a_j}$. We define the merging polynomial of $T$ (written $\nodepol_\JJ(T)$) as the determinant of the resulting matrix.
\end{defn}
Note that the merging polynomial $\nodepol(T)$ is a polynomial in $x_{a_1}, x_{a_2}, \ldots$. For example, consider the algorithm's execution from Fig. \ref{fig:tree_algo_2}. At stage 2 of the algorithm, the merging tree $T$ with root $\{4,32,40,48,56\}$ has a square system of equations corresponding to the matrix\\
 {
 \resizebox{0.8\columnwidth}{!}{%
$\hspace{3cm}
\nodemat(T) = \begin{pmatrix}
1 & 1 & 0 & 0 \\
0 & 0 & 1 & 1 \\
\exp(2 \pi i 40 /N) & \exp(2 \pi i 56 /N) & \exp(2 \pi i 32 /N) & \exp(2 \pi i 48 /N)\\
\exp(2 \pi i (2*40) /N) & \exp(2 \pi i (2*56) /N) & \exp(2 \pi i (2*32) /N) & \exp(2 \pi i (2*48) /N)
\end{pmatrix}
$}\\
}

Note that the lead node labels are \(a_1 = 40, a_2 = 56, a_3 = 32, a_4= 48\); and the entries in column $j$ of this matrix are zero or powers of $\exp(2 \pi i a_j /N)$. By replacing these roots of unity as above, we get
\begin{align*}
\nodepol(T) &= \det \begin{pmatrix}
1 & 1 & 0 & 0 \\
0 & 0 & 1 & 1 \\
x_{a_1} & x_{a_2} & x_{a_3} & x_{a_4} \\
x_{a_1}^2 & x_{a_2}^2 & x_{a_3}^2 & x_{a_4}^2
\end{pmatrix} \\ &= (x_{a_1}-x_{a_2}) (x_{a_3}-x_{a_4}) (x_{a_1}+x_{a_2}-x_{a_3}-x_{a_4})
\end{align*}
The number of variables in the merging polynomial $\nodepol(T)$ is $w_\JJ(T)$.

\begin{lemma}
\label{lem:poly_merging}
The merging polynomial $\nodepol_\JJ(T)$  is a non-zero polynomial with a degree at most $w_\JJ^2(T)$.
\end{lemma}
\begin{IEEEproof}
We first show that $\nodepol_\JJ(T)$ is a non-zero polynomial. For this, note that if $\JJ$ is a singleton, then the merging tree is simply a single node (the algorithm concludes in the first stage). In this case, the merging polynomial is $1$, hence non-zero.

For an arbitrary $\JJ$, let $T$ be a complete merging tree in $\TT(\JJ)$. As discussed previously, let $\JJ'$ be obtained by removing an element $a_{11}$ from the leftmost leaf of $T$ from  $\JJ$. Consider taking the Laplace expansion of the determinant. The coefficient of the highest power of $x_{a_{11}}$ is given by the cofactor obtained by removing the first column and last row from the corresponding matrix. By a direct application of Lemma \ref{lem:structure_merg_matrix}, this coefficient is given by 
\(
\prod_{i} \nodepol_{\JJ'}(T_i).
\)
The non-zeroness of $\nodepol$ follows from induction on the size of $\JJ$.

For the degree, note that $\nodepol$ is a determinant of size $w_\JJ(T)$. The $i^\textsf{th}$ row in the determinant has a degree at most $i$. Since the terms in the determinant involve exactly one element from each row, the degree of each term in the determinant is at most $\sum_{i=1}^{w_\JJ(T)} i \leq w_\JJ^2(T)$.
\end{IEEEproof}


\begin{lemma}
\label{lem:prob_inv_merg_matrix}
Suppose $k = O(N^\alpha)$ for some $\alpha <1$. For any complete merging tree $T$ the merging matrix $\nodemat(T)$ is invertible with probability $1-O(\log^2N/N^{1-\alpha})$. 
\end{lemma}

\begin{IEEEproof}
Consider a merging tree $T$ with $x$ leaf nodes $a_1, a_2, \ldots, a_x$. Let the elements of leaf $a_j$ be $a_{j1}, a_{j2}, \ldots$. As before, we let $\mu_j = \mu_\JJ(a_j) = |\nodelab(a_j)|$ be the number of elements in leaf $a_j$. 

We note that the determinant of the merging matrix $\nodemat(T)$ is obtained by evaluating the merging polynomial $\nodepol(T)$ at $x_{a_{jk}} = \exp( 2 \pi i a_{jk} /N)$. Thus, the merging matrix is invertible iff the merging polynomial $\nodepol(T)$ does not have $x_{a_{jk}}=\exp(2 \pi i a_{jk}/N)$ as a root. Since our support $\JJ$ is picked randomly, we apply the Schwartz-Zippel lemma \cite{zippel1979probabilistic,demillo1977probabilistic} to show that the merging polynomial is unlikely to vanish at most choices of the support.

 If $\nodelab(a_j)$ be the label of the leaf $a_j$, recall that $\nodelab_j \coloneqq \nodelab_{\ZZ_N}(a_j)$ is the set of all possible indices at leaf $a_j$; and we have $\nodelab_\JJ(a_j)= \nodelab_j \cap \JJ $. Next, we discuss the probability of the merging polynomial vanishing for the probability model on $\JJ$ discussed earlier.

Before we proceed, we need the following notation: For a set $A$, we define $A^{\vec{\mu}}$ to be the set of all unordered $\mu-$ tuples with each element in the tuple picked from $A$ without repetition.

Thus the elements from leaf $a_j$ are drawn from $\nodelab_j^{\vec{\mu}_j}$. If we denote by $n$ the size of $\nodelab_j$, i.e. $n=|\nodelab_j| = N/2^r$, then there are $\binom{n}{\mu_j}$ possible choices for the leaf elements of $a_j$. Note that all these choices have the same merging pattern; moreover, based on our probability model, all these choices have equal probability.

So we define
\[S = \exp\left(\frac{2 \pi i}{N} \left(\nodelab_1^{\vec{\mu}_1} \times \nodelab_2^{\vec{\mu}_2} \times \dots \right)\right), \text{ where }\nodelab_i = \nodelab_{\ZZ_N}(l_i).\]
Note that $S$ is a finite subset of $\mathbb{C}^{w_\JJ(T)}$, the domain of $\nodepol_\JJ(T)$, and (according to the probability model on $\JJ$) all choices in $S$ are equally likely. Now consider the polynomial $\nodepol_\JJ$ evaluated on $S$: applying the Schwarz Zippel Lemma, we get that the probability of the polynomial $\nodepol(T)$ vanishing is bounded by 
\[
\Prob\left(\text{merging matrix }\nodemat_T\text{ non-invertible} \right) \leq \text{deg}(\nodepol_\JJ(T))/|S| \]\[\leq \frac{1}{|S|}w_\JJ^2(T) = \frac{1}{\prod_j \binom{n}{\mu_j}}\left(\sum_{j} \mu_j \right)^2.
\]
Now, since $\mu_j$ are leaf sizes in a complete merging tree, from Lemma 2(6), we have $\eta < \mu_j \leq \eta (r+1)$. Combining this with $\binom{n}{\mu_j} \geq n$ for $0<\mu_j<n$, we get

\begin{equation}
\begin{split}
\Prob\left(\text{merging matrix }\nodemat_T\text{ non-invertible} \right) \leq \frac{\eta^2 (r+1)^2 x^2}{n^x} \\ \leq \frac{\eta^2 (r+1)^2}{n}= \frac{\eta^2 (r+1)^2 2^r}{N} 
\leq \frac{8\eta^2 k \log^2 k }{N},
\end{split}
\end{equation}
where in the second inequality we used that $x^2/n^x \leq 1/n $ for any $x\geq 1, n>4$; and in the final inequality we use $r \leq 2\log k$ and $2^r \leq 2k$ (recall that $r = \lceil \log k \rceil$).
For any $k = O(N^{\alpha})$, where $0<\alpha<1$ probability that merging matrix $\nodemat_T$ is not invertible  is
\[
(\text{const }) (1-\alpha)^2 \frac{\log^2 N}{N^{1-\alpha}} \rightarrow 0.
\]

\end{IEEEproof}

\subsection{Complexity}

The computations done by the algorithm can be seen to have these three components: 1) Computing the DFTs of downsampled signals at each stage of the algorithm, 2) Subtracting the effect of known coefficients (coefficients identified in the previous stages) from the computed DFT values, and 3) Solving system of equations at roots of complete merging trees. In the sequel, we will bound each component's required computations.

\begin{enumerate}
    \item At stage $i$ of the algorithm, we see that the size of the downsampled signal is $2^{r-i}$. Note that there are $\neqlvl$ DFTs to be computed at each stage. Recall that each DFT of size $N$ takes $c_1N \log N$ computations. Thus, the overall complexity of computing these DFTs is bounded by
 \[\neqlvl c_1 \sum_{i=0}^{r}2^{r-i} \log 2^{r-i} = \neqlvl c_1 \sum_{i=0}^{r}(r-i)2^{r-i} = \neqlvl c_1 \left(2^{r+1}(r-1)+2\right)= O(k \log k)\]    
\item At each stage of the algorithm, once the DFT coefficients are computed, the contribution of known DFT coefficients needs to be subtracted. Note that each leaf node contributes to at most one DFT coefficient at any stage of the algorithm. The value to be subtracted is the induced weight of the resolved leaves. The total number of coefficients is $k$, and so the number of known coefficients (resolved leaves) is at most $k$.  Thus, this subtraction can be done in $O(k)$.

The maximum number of stages in the algorithm is $r$. The overall complexity for this component is $O(kr)=O(k \log k)$.

\item From Lemma \ref{lem:system_size}, for any leaf, the expected size of the system of equations from leaf to root is $O(1)$. There are $2^r$ such leaves; thus, the total average size of the system of equations is bounded by $2^r O(1)$, which is equal to $O(k)$.

\end{enumerate}

\begin{lemma}
\label{lem:system_size}
For any leaf $l$ of $\TT_r(\JJ)$, the expected size of the system of equations involving unknowns from $l$ is $O(1)$.
\end{lemma}
\begin{IEEEproof}
Consider the path from leaf $l$ to the root of $\TT_r(\JJ)$. For each node $i$ in the path, from $i=0$ at the leaf $l$ to $i=r$ at the root, consider $n_i$ to be the number of unknowns that merge in this path at node $i$ ($n_i=|\nodelab(i)|$).
Thus, $n_i$ is the number of unknowns contributed by the subtree of $\TT_r$ rooted at $j$. For $n_i$ to be at least one, a merging tree must extend up to $j$, with non-zero skewness at $j$.

The complexity of solving the systems of equations for each leaf node is bounded by $c_2\sum_{i =0}^r n_i^3$. The expected value of this complexity is 
\[c_2E(\sum_{i =0}^r n_i^3) =  c_2 \sum_{i =0}^r E(n_i^3)\leq \sum_{i =0}^\infty e^{\frac{\eta+109}{6}}  e^{-\frac{i(\eta-18 \ln 2)}{6}} = O(1).\]
\end{IEEEproof}

\section{Final result}

\begin{theorem}
\label{thm:correctness_algo_2}
Suppose $k= O(N^\alpha)$ for some $0<\alpha<1$. Then, for $\JJ \sim \Z_N^{\downarrow k}$, the proposed shift and progressive sample algorithm is a $(1,k\log k)$ SoE-SDFT algorithm\footnote{The complexity and size of system of equations are computed as an average over $\JJ \sim \Z_N^{\downarrow k}$. The algorithm succeeds with probaility $1-o(1)$ for $\JJ \sim \Z_N$}.  
\end{theorem}
\begin{IEEEproof}
To prove correctness, we proved the following:
\begin{enumerate}
    \item Every leaf $i$ of $\TT_r(\JJ)$ is part of a complete merging tree with high probability; and
    \item Every complete merging tree has an invertible merging matrix with high probability.
\end{enumerate}
For 1), note that for a given leaf node $i$, the probability that it is not part of a complete merging tree is  $O(1/k^{\eta/3})$ from Lemma \ref{lem:prob_height_unresolved} and corollary \ref{cor:prob_unresolved_merg_tree}. By union bound, the probability that some leaf is not part of a complete merging tree is $O(1/k^{\eta/3-1})$. This probability goes to zero for $\eta>4$ as $N \rightarrow \infty$.

For 2), note that if some complete merging matrix $T$ has a non-invertible merging matrix $\nodemat(T)$, then the algorithm reports a failure, and all nodes on the path from $\text{root(T)}$ to the root of $\TT_r(\JJ)$ cannot have an invertible merging matrix. Consider an ancestor of $\text{root(T)}$, say $a$ with a square merging matrix to see this. Then $\nodemat(a)$ has the form \\
\resizebox{0.6\columnwidth}{!}{%
$\hspace{3cm}
\nodemat(a) = \begin{pNiceArray}{cccccccc}[first-row,first-col]
  &  &   & & &   & & &       \\
  &  & \Block[draw,rounded-corners]{3-3}{\nodemat(T)}  &  &  & \Block{3-3}<\huge>{0}  & & &   \\
    &   &  &  &   & &  &   &  \\
  &  &   &  &  &   & & &    \\
  &  & \Block{3-3}<\huge>{*}  &  &  & \Block[draw,rounded-corners]{3-3}{M}  & &  \\
    &   &  &  &   & &  &   &   \\
  &  &   &  &  &   & & &     
\end{pNiceArray}
$}\\
So that $\det \nodemat(a) = \det \nodemat(T)\times \det M$, and since $\nodemat(T)$ is nonivertible, $\nodemat(a)$ is also non invertible. In particular, $\nodemat(\text{root})$ is either non-invertible or wide. But from Lemma \ref{lem:prob_inv_merg_matrix}, the probability that a node matrix is invertible is $O(\log^2N/N^{1-\alpha})$.

Thus, the probability of the algorithm failing is bounded by 
\[
(\text{const})\frac{1}{k^{\eta/3-1}} + (\text{const})\frac{\log^2 N}{N^{1-\alpha}} \rightarrow 0.
\]
\end{IEEEproof}

\section{Simulations}

\begin{figure*}[ht]
\centering
\subfloat[]{\includegraphics[width=0.25\textwidth]{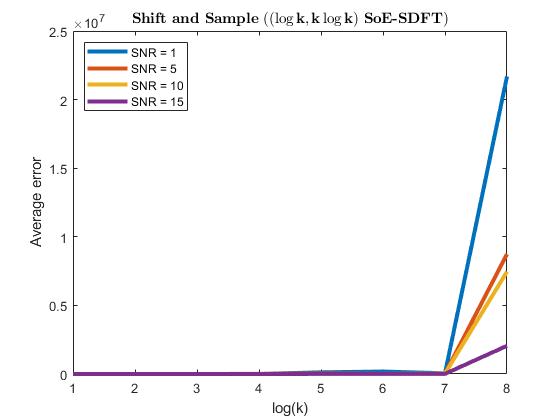}}
\subfloat[]{\includegraphics[width=0.25\textwidth]{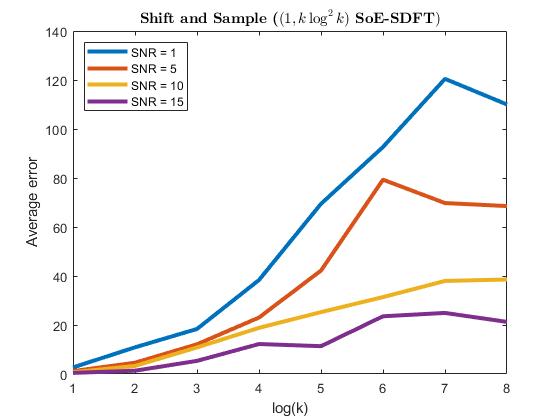}}
\subfloat[]{\includegraphics[width=0.25\textwidth]{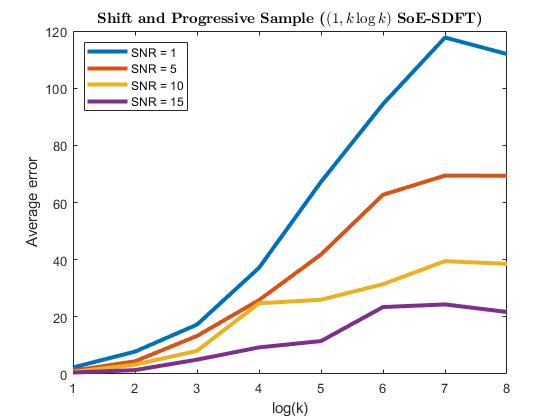}}
\caption{\scriptsize The Average error of algorithms  w.r.t $\log k$. The plot is obtained for four SNR regimes: a) Shift and sample algorithm upper bound, b) Shift and sample algorithm lower bound (note that here the complexity is $O(k \log^2 k)$), c) Shift and progressive sample (proposed) algorithm. Each point on the plot is obtained by averaging over 10000 runs and the length of the signal $N = 2^{14}$.}
\label{fig:error_vs_logk}
\end{figure*}

\begin{figure*}[ht]
\centering
\subfloat[]{\includegraphics[width=0.25\textwidth]{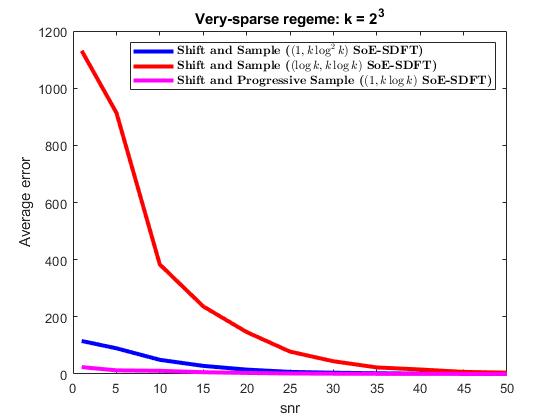}}
\subfloat[]{\includegraphics[width=0.25\textwidth]{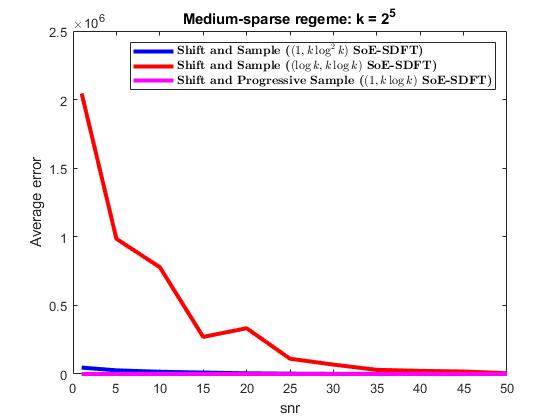}}
\subfloat[]{\includegraphics[width=0.25\textwidth]{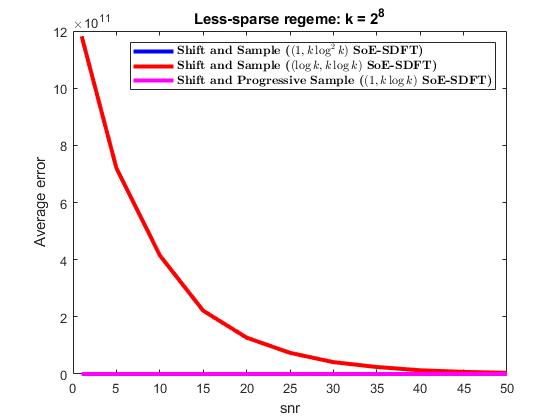}}
\caption{\scriptsize The Average error of algorithms  w.r.t SNR. Here, the length of the discrete signal is $N = 2^{14}$. The plot is obtained for three different sparsity regimes: a) Very-sparse regime: $k = 2^3$, b) Medium-sparse regime: $k = 2^5$, c) Less-sparse regime: $k = 2^{8}$. Each point on the plot is obtained by averaging over 10000 runs.}
\label{fig:error_vs_snr}
\end{figure*}
\begin{enumerate}
    \item \emph{Computation model:} We investigate the algorithm's stability to noisy computations: in particular, we assume that the input to every system of equation sub-block in the SoE-SDFT algorithm is corrupted with white Gaussian noise.
    \item \emph{Comparision methods:} We compare the stability of the proposed algorithm to $(\log k, k \log k)$ SoE-SDFT algorithm and $(1, k \log^2 k)$ SoE-SDFT algorithm obtained from shift and sample algorithm \cite{10308632} (also mentioned in Section \ref{sec:shift-and-sample}). Since the system of equations is $O(\log k)$ for the $(\log k, k \log k)$ SoE-SDFT algorithm, the stability is expected to be worse. 
    \item \emph{Comparision metric:} We plot the average $l_2$ norm of the difference between original and estimated DFT coefficients as a function of $k$ at various SNR regimes and vice versa.
\end{enumerate}
\begin{figure*}[ht]
\centering
\subfloat[]{\includegraphics[width=0.33\textwidth]{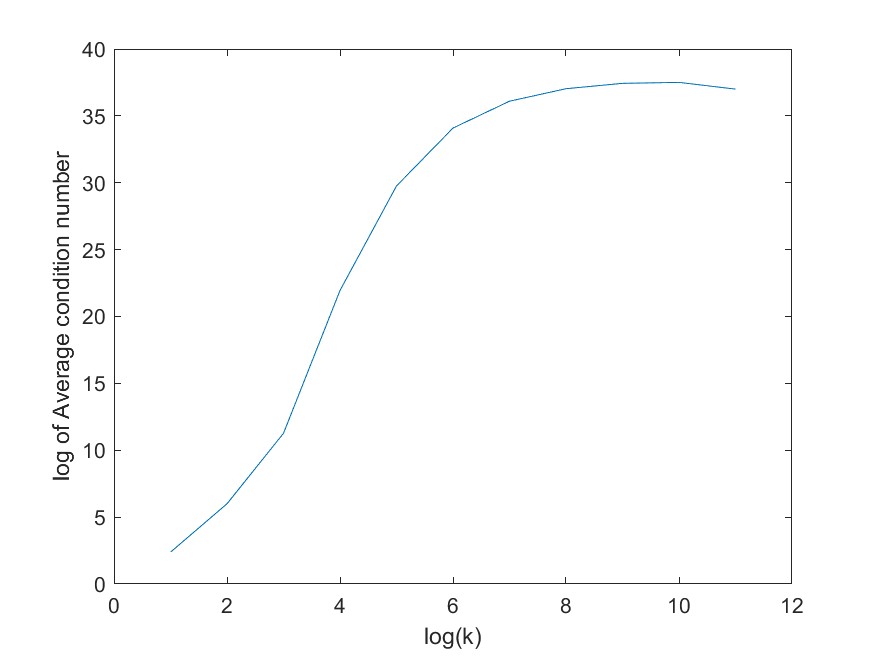}}
\subfloat[]{\includegraphics[width=0.33\textwidth]{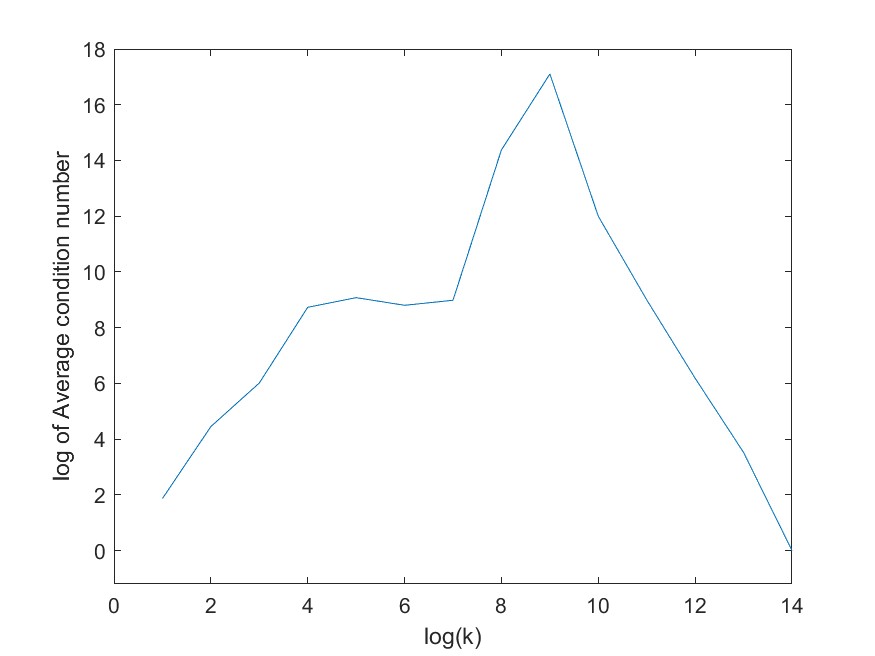}}
\subfloat[]{\includegraphics[width=0.33\textwidth]{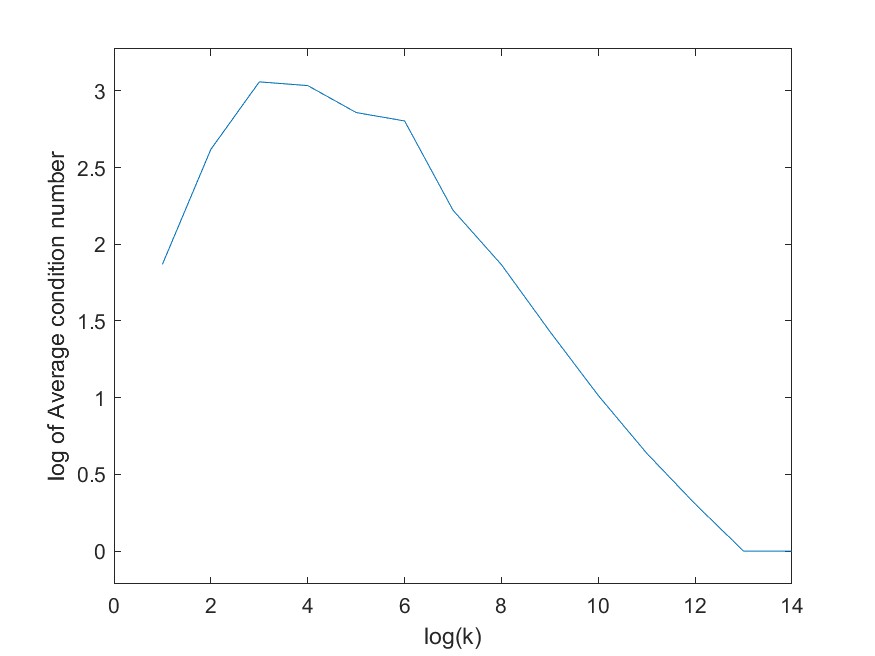}}
\caption{\scriptsize The Average condition number (log scale) of random DFT submatrices (created by picking first $k$ rows and $k$ random columns), Shift and Sample algorithm, and Shift and Progressive Sample algorithm w.r.t $\log k$. Here, the length of the discrete signal is $N = 2^{14}$. Each point on the plot is obtained by averaging over 10000 runs.}
\label{fig:sim_cond}
\end{figure*}

\section{List of tools/existing results used}
 We use the following tools in analyzing the proposed algorithms

\begin{enumerate}




\item \emph{(Chernoff Upper Tail \cite{harvey2022randomized})}Let $X_1,X_2,\ldots$ be i.i.d random variables such that $X_i\in[0,1]$. Define sum of these i.i.d random variables $X = \sum_i X_i$ and $E(X) = \sum_i E(X_i)$. For any $\delta > 1$, 
\begin{equation}
\label{eq:chernoff_upper}
\Prob \left(X \geq (1+ \delta)E(X)\right) \leq \exp \left(-\delta E(X)/3)\right).
\end{equation}


\item \emph{(Cauchy-Schwarz inequality \cite{mukhopadhyay2020probability})} For any two random variables $X$ and $Y$, 

\begin{equation}
\label{eq:c_s_ineq}
(E(XY))^2 \leq E(X^2) E(Y^2).
\end{equation}

\item\emph{(Schwartz-Zippel Lemma)}\cite{zippel1979probabilistic,demillo1977probabilistic}) Let $p(x_1,x_2,\ldots)$ be a polynomial of total degree $d$. Assume that $p$ is not identically zero. Let $S\subseteq \mathbb{C}$ be any finite set. Then, if we pick $y_1,y_2,\ldots$ independently and uniformly from $S$,
\begin{equation}
\label{eq:schwartz_zippel}
\Prob[p(y_1,y_2,\ldots) = 0]\leq\frac{d}{|S|}.
\end{equation}
\item \emph{(Bound on moments of binomial distribution)}\cite{berend2010improved,ahle2022sharp} Let $X$ be a binomial random variable with mean $E(X)$. Then, moments of $X$ are bounded by
\begin{equation}
\label{eq:bnd-sub-poisson}
E(X^j) \leq E(X)^j \exp(j^2/2E(X)).
\end{equation}


\begin{figure}[ht]
\centering
\resizebox{0.3\columnwidth}{!}{%
\begin{tikzpicture}[scale=0.5]
\node (0) at (0,8) [ minimum size = 0.25cm, fill=gray!10, draw] {$1$};
\node (1.1) at (-4,5) [ minimum size = 0.25cm, fill=gray!10, draw] {$2$};
\node (1.2) at (4,5) [ minimum size = 0.25cm, fill=gray!10, draw] {$3$};
\node (2.1) at (-6,2) [ minimum size = 0.25cm, fill=gray!10, draw] {$4$};
\node (2.2) at (-2,2) [ minimum size = 0.25cm, fill=gray!10, draw] {$5$};
\node (2.3) at (2,2) [ minimum size = 0.25cm, fill=gray!10, draw] {$6$};
\node (2.4) at (6,2) [ minimum size = 0.25cm, fill=gray!10, draw] {$7$};

\draw (node cs:name=0) --node[above left] {} (node cs:name =1.1);
\draw (node cs:name=0) -- node[above right] {}(node cs:name =1.2);

\draw (node cs:name=1.1) --node[above left] {} (node cs:name =2.1);
\draw (node cs:name=1.1) --node[above right] {} (node cs:name =2.2);
\draw (node cs:name=1.2) --node[above left] {} (node cs:name =2.3);
\draw (node cs:name=1.2) --node[above right] {} (node cs:name =2.4);

\end{tikzpicture}}
\caption{\scriptsize Post order traversal: $4 \rightarrow 5 \rightarrow 2 \rightarrow 6 \rightarrow 7 \rightarrow 3 \rightarrow 1$.}
\label{fig:post_order_traversal}
\end{figure}
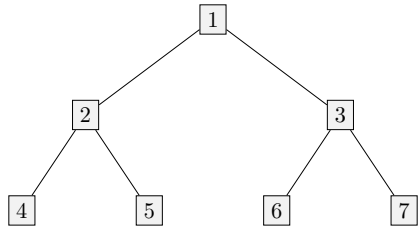
\item \emph{(Post-order traversal)} In some of our proofs, we find it useful to list all the nodes in congruence trees by post-order traversal (left-right-root) \cite{knuth1973art}. For any congruence tree $T$, we refer by nodes($T$) to the list of all nodes ordered by a post-order traversal of $T$. In a post-order traversal, the left nodes in any subtree are listed first, followed by the right nodes, and then the root. In particular, we recall the following: node $a$ appears earlier than node $b$ in nodes($T$), if and only if either there exists a subtree with $a$ on the left and $b$ on the right or $b$ itself is the root of a subtree containing $a$ on the left. 
\end{enumerate}




\begin{thebibliography}{10}
	\providecommand{\url}[1]{#1}
	\csname url@samestyle\endcsname
	\providecommand{\newblock}{\relax}
	\providecommand{\bibinfo}[2]{#2}
	\providecommand{\BIBentrySTDinterwordspacing}{\spaceskip=0pt\relax}
	\providecommand{\BIBentryALTinterwordstretchfactor}{4}
	\providecommand{\BIBentryALTinterwordspacing}{\spaceskip=\fontdimen2\font plus
		\BIBentryALTinterwordstretchfactor\fontdimen3\font minus \fontdimen4\font\relax}
	\providecommand{\BIBforeignlanguage}[2]{{%
			\expandafter\ifx\csname l@#1\endcsname\relax
			\typeout{** WARNING: IEEEtran.bst: No hyphenation pattern has been}%
			\typeout{** loaded for the language `#1'. Using the pattern for}%
			\typeout{** the default language instead.}%
			\else
			\language=\csname l@#1\endcsname
			\fi
			#2}}
	\providecommand{\BIBdecl}{\relax}
	\BIBdecl
	
	\bibitem{donoho-stark:uncertainty}
	D.~Donoho and P.~Stark, ``Uncertainty principles and signal recovery,'' \emph{{SIAM} J. Appl. Math.}, vol.~49, no.~3, pp. 906--931, 1989.
	
	\bibitem{DS-1}
	B.~Osgood, A.~Siripuram, and W.~Wu, ``Discrete sampling and interpolation: Universal sampling sets for discrete bandlimited spaces,'' \emph{IEEE Trans. Information Theory}, vol.~58, no.~7, pp. 4176--4200, 2012.
	
	\bibitem{parker1964inverses}
	F.~Parker, ``Inverses of vandermonde matrices,'' \emph{The American Mathematical Monthly}, vol.~71, no.~4, pp. 410--411, 1964.
	
	\bibitem{gohberg1997fast}
	I.~Gohberg and V.~Olshevsky, ``The fast generalized parker--traub algorithm for inversion of vandermonde and related matrices,'' \emph{Journal of Complexity}, vol.~13, no.~2, pp. 208--234, 1997.
	
	\bibitem{10308632}
	C.~R. Pochimireddy, A.~Siripuram, and B.~Osgood, ``Fast dft computation for signals with structured support,'' \emph{IEEE Transactions on Information Theory}, pp. 1--1, 2023.
	
	\bibitem{osgood2018lectures}
	B.~Osgood, \emph{Lectures on the Fourier Transform and Its Applications}.\hskip 1em plus 0.5em minus 0.4em\relax American Mathematical Society, 2018.
	
	\bibitem{cooley1965algorithm}
	J.~W. Cooley and J.~W. Tukey, ``An algorithm for the machine calculation of complex fourier series,'' \emph{Mathematics of computation}, vol.~19, no.~90, pp. 297--301, 1965.
	
	\bibitem{good1958interaction}
	I.~J. Good, ``The interaction algorithm and practical fourier analysis,'' \emph{Journal of the Royal Statistical Society. Series B (Methodological)}, pp. 361--372, 1958.
	
	\bibitem{rader1968discrete}
	C.~M. Rader, ``Discrete fourier transforms when the number of data samples is prime,'' \emph{Proceedings of the IEEE}, vol.~56, no.~6, pp. 1107--1108, 1968.
	
	\bibitem{Pan2016}
	\BIBentryALTinterwordspacing
	V.~Y. Pan, ``How bad are vandermonde matrices?'' \emph{SIAM Journal on Matrix Analysis and Applications}, vol.~37, no.~2, pp. 676--694, 2016. [Online]. Available: \url{https://doi.org/10.1137/15M1030170}
	\BIBentrySTDinterwordspacing
	
	\bibitem{3055462}
	V.~{Cevher}, M.~{Kapralov}, J.~{Scarlett}, and A.~{Zandieh}, ``An adaptive sublinear-time block sparse fourier transform,'' \emph{Proceedings of the 49th Annual ACM SIGACT Symposium on Theory of Computing}, p. 702–715, 2017.
	
	\bibitem{gilbert2014recent}
	A.~C. Gilbert, P.~Indyk, M.~Iwen, and L.~Schmidt, ``Recent developments in the sparse fourier transform: A compressed fourier transform for big data,'' \emph{IEEE Signal Processing Magazine}, vol.~31, no.~5, pp. 91--100, 2014.
	
	\bibitem{Kap17}
	\BIBentryALTinterwordspacing
	M.~Kapralov, ``Sample efficient estimation and recovery in sparse fft via isolation on average,'' in \emph{2017 IEEE 58th Annual Symposium on Foundations of Computer Science (FOCS)}.\hskip 1em plus 0.5em minus 0.4em\relax Los Alamitos, CA, USA: IEEE Computer Society, oct 2017, pp. 651--662. [Online]. Available: \url{https://doi.ieeecomputersociety.org/10.1109/FOCS.2017.66}
	\BIBentrySTDinterwordspacing
	
	\bibitem{GSS22}
	Y.~Gao, Z.~Song, and B.~Sun, ``An $o(k\log n)$ time fourier set query algorithm,'' 2022.
	
	\bibitem{SSWZ22}
	Z.~Song, B.~Sun, O.~Weinstein, and R.~Zhang, ``Sparse fourier transform over lattices: A unified approach to signal reconstruction,'' 2022.
	
	\bibitem{candes:robust}
	E.~Candes, J.~Romberg, and T.~Tao, ``Robust uncertainty principles: exact signal reconstruction from highly incomplete frequency information,'' \emph{IEEE Transactions on Information Theory}, vol. 52, (2), pp. 489-- 509, 2006.
	
	\bibitem{pawar-ramachandran}
	S.~Pawar and K.~Ramchandran, ``A ffast framework for computing a k-sparse dft in o(k log k) time using sparse-graph alias codes,'' \emph{IEEE International Symposium on Information Theory}, 2015.
	
	\bibitem{ghazi2013sample}
	B.~Ghazi, H.~Hassanieh, P.~Indyk, D.~Katabi, E.~Price, and L.~Shi, ``Sample-optimal average-case sparse fourier transform in two dimensions,'' in \emph{2013 51st Annual Allerton Conference on Communication, Control, and Computing (Allerton)}.\hskip 1em plus 0.5em minus 0.4em\relax IEEE, 2013, pp. 1258--1265.
	
	\bibitem{kapralov2019dimension}
	M.~Kapralov, A.~Velingker, and A.~Zandieh, ``Dimension-independent sparse fourier transform,'' in \emph{Proceedings of the Thirtieth Annual ACM-SIAM Symposium on Discrete Algorithms}.\hskip 1em plus 0.5em minus 0.4em\relax SIAM, 2019, pp. 2709--2728.
	
	\bibitem{mukhopadhyay2020probability}
	N.~Mukhopadhyay, \emph{Probability and statistical inference}.\hskip 1em plus 0.5em minus 0.4em\relax CRC Press, 2020.
	
	\bibitem{nearlyoptimalsparse}
	H.~{ Hassanieh}, P.~{Indyk}, D.~{Katabi}, and E.~{Price}, ``Nearly optimal sparse fourier transform,'' \emph{44th Symposium on Theory of Computing}, p. 563–578, 2012.
	
	\bibitem{berend2010improved}
	D.~Berend and T.~Tassa, ``Improved bounds on bell numbers and on moments of sums of random variables,'' \emph{Probability and Mathematical Statistics}, vol.~30, no.~2, pp. 185--205, 2010.
	
	\bibitem{ahle2022sharp}
	T.~D. Ahle, ``Sharp and simple bounds for the raw moments of the binomial and poisson distributions,'' \emph{Statistics \& Probability Letters}, vol. 182, p. 109306, 2022.
	
	\bibitem{zippel1979probabilistic}
	R.~Zippel, ``Probabilistic algorithms for sparse polynomials,'' in \emph{International symposium on symbolic and algebraic manipulation}.\hskip 1em plus 0.5em minus 0.4em\relax Springer, 1979, pp. 216--226.
	
	\bibitem{demillo1977probabilistic}
	R.~A. DeMillo and R.~J. Lipton, ``A probabilistic remark on algebraic program testing.'' GEORGIA INST OF TECH ATLANTA SCHOOL OF INFORMATION AND COMPUTER SCIENCE, Tech. Rep., 1977.
	
	\bibitem{harvey2022randomized}
	N.~Harvey, ``A first course in randomized algorithms,'' 2022.
	
	\bibitem{knuth1973art}
	D.~E. Knuth \emph{et~al.}, \emph{The art of computer programming}.\hskip 1em plus 0.5em minus 0.4em\relax Addison-Wesley Reading, MA, 1973, vol.~3.
	
\end{thebibliography}
\end{document}